\documentclass[aap]{imsart}
\usepackage{amsfonts}
\usepackage{amssymb}
\usepackage{amstext}
\usepackage{amsmath, amsthm}
\usepackage{verbatim}
\usepackage{xspace}
\usepackage{graphicx}
\usepackage{algorithm}
\usepackage{amsmath}
 \usepackage{amsfonts}
 \usepackage{amssymb}
  \usepackage[dvipsnames,usenames]{color}
\newtheorem{theorem}{Theorem}

\newtheorem{lemma}{Lemma}
\newtheorem{corollary}[theorem]{Corollary}

\newtheorem{definition}{Definition}

\newtheorem{fact}{Fact}

\makeatletter
\def\newxxxproof#1{\@nxxxprf{#1}}

\def\@nxxxprf#1#2{\@xnxxxprf{#1}{#2}}

\def\@xnxxxprf#1#2{\expandafter\@ifdefinable\csname #1\endcsname
\global\@namedef{#1}{\@xxxprf{#1}{#2}}\global\@namedef{end#1}{\@endxxxproof}}

\def\@xxxprf#1#2{\@xxxxprf{#1}{#2}}

\def\@xxxxprf#1#2{\@beginxxxproof{#2}{\csname the#1\endcsname}\ignorespaces}

\def\@beginxxxproof#1{\rm \trivlist \item[\hskip \labelsep{\sc #1.\/}]}
\def\@endxxxproof{\outerparskip 0pt\endtrivlist}

\newxxxproof{@xxxproof}{Main Theorem}

\makeatother

\newcommand{\one}{\mathbf{1}}

\newcommand{\lab}{\label}

\def\beq{\begin{eqnarray}}
\def\eeq{\end{eqnarray}}
\def\ben{\begin{enumerate}}
\def\een{\end{enumerate}}
\def\beqs{\begin{eqnarray*}}
\def\eeqs{\end{eqnarray*}}
\def\bel{\begin{lemma}}
\def\eel{\end{lemma}}
\def\bed{\begin{definition}}
\def\eed{\end{definition}}
\newcommand{\F}{\breve{\tilde{F}}}
\newcommand{\X}{X}
\newcommand{\Y}{Y}

\newcommand{\R}{\mathbb{R}}
\newcommand{\RR}{\mathbb{R}}
\newcommand{\C}{\mathit{\beta_{fat}}}

\newcommand{\M}{\mathcal{M}}

\newcommand{\Tr}{\mathbf{Tr}}

\newcommand{\aaa}{\alpha}

\newcommand{\eps}{\epsilon}

\newcommand{\E}{\mathbb{E}}
\newcommand{\p}{\mathbb{P}}

\newcommand{\PP}{\mathcal{P}}

\newcommand{\LL}{\mathcal L}
\newcommand{\MM}{\mathcal M}
\newcommand{\NN}{\mathcal N}
\newcommand{\ra}{\rightarrow}
\newcommand{\la}{\lambda}

\def\ie{i.\,e.\,}
\def\wp{with probability\,}
\def\wpot{with probability $\geq 1- 10^{-3}$\,}
\def\byfactone{by Fact~\ref{1f1}}
\def\ot{1- 10^{-3}\,}

\newcommand{\de}{\delta}

\newcommand{\vol}{\mathrm{vol}}
\newcommand{\De}{\Delta}
\newcommand{\rr}{\mathbf{r}}
\renewcommand{\a}{\alpha}

\newcommand{\tf}{\hat{F}}

\newcommand{\tg}{\hat{G}}
\newcommand{\tv}{\hat{V}}
\newcommand{\tk}{\hat{K}}
\renewcommand{\l}{\lambda}
\newcommand{\xx}{W}
\newcommand{\yy}{Z}
\newcommand{\AAA}{A}
\newcommand{\BBB}{B}
\newcommand{\dm}{d_{\mathcal M}}
\newcommand{\mA}{\mathbf{A}}

\begin{document}

\title{Randomized Interior Point Methods for Sampling and Optimization
}


\author{Hariharan Narayanan \\
{ Department of Statistics and Department of Mathematics\\
University of Washington, Seattle, 98105, USA\\
harin@uw.edu}}


%

\begin{abstract}
 We present a Markov Chain, ``Dikin walk", for sampling from a convex body equipped with a self-concordant barrier. This Markov Chain corresponds to a natural random walk with respect to a Riemannian metric defined using the Hessian of the barrier function.

 For every convex set of dimension $n$, there exists a self-concordant barrier whose self-concordance parameter is $O(n)$.
Consequently, a rapidly mixing Markov Chain of the kind we describe can be defined (but not always be efficiently implemented) on any convex set.
We  use these results to design an  algorithm consisting of a single random walk for optimizing a linear function on a convex set.
Using results of Barthe \cite{barthe} and Bobkov and Houdr\'{e} \cite{bobkov}, on the isoperimetry of products of weighted Riemannian manifolds, we obtain sharper upper bounds on the mixing time of a Dikin walk on products of convex sets than the bounds obtained from a direct application of the Localization Lemma.
The results in this paper generalize previous results of \cite{KNsample} from polytopes to spectrahedra and beyond, and improve upon those results in a special case when the convex set is a direct product of lower dimensional convex sets.
This Markov Chain like the chain described in \cite{KNsample} is affine-invariant.
\end{abstract}
\maketitle

\section{Introduction}
The task of sampling from distributions supported in high dimensional Euclidean space arises frequently in Statistics. In particular, the question of sampling from a nearly uniform distribution supported on  high dimensional convex set arises naturally in the task of computing the  volume \cite{DFK} of the latter.
Markov Chains that are rapidly mixing are a tool for sampling.
The usual strategy for sampling from distributions supported on high dimensional Euclidean space is to design a rapidly mixing Markov Chain whose stationary distribution is the desired distribution, run it for sufficiently long and then pick the final point as a sample.

Previous sampling algorithms were applicable to convex sets
specified in the following way. The input consists of an $n$-dimensional
convex set $K$ circumscribed around, and inscribed in, balls of radius
$r$ and $R$ respectively. The algorithm has access to an oracle which
when supplied with a point in $\R^n$ answers ``yes" if the point is
in $K$ and ``no" otherwise. If these balls are not provided, the convex set could be either extremely small or extremely large, making it impossible to sample the convex set in any fixed time frame.

The first polynomial time algorithm for sampling convex sets
appeared in \cite{DFK}. It did a random walk on a sufficiently dense
grid. The dependence of its mixing time on the dimension was
$O^*(n^{23})$. It resulted in the first randomized polynomial time
algorithm to approximate the volume of a convex set.
A long series of works directed to improving the complexity culminated in the work of Lov\'{a}sz and Vempala,
who achieve a mixing time of $O^*(n^4)$ in \cite{LV}.
Another random walk for sampling convex sets, known as the ball walk,
 does the following. Suppose the
current point is $x_i$. A point $y$ is chosen uniformly at random from a
ball of radius $\delta$ centered at $x_i$. If $y \in K$, $x_{i+1}$
is set to $y$; otherwise $x_{i+1} = x_i.$ After many successive
improvements over several papers, it was shown in \cite{KLS} that  if $\delta < \frac{r}{\sqrt{n}}$
a ball walk mixes in $O^*(n \frac{R^2}{\delta^2})$ steps from a warm
{start (\ie a distribution whose density with respect to the uniform measure is bounded above by a universal constant).  
A third  random walk analyzed more
recently by Lov\'{a}sz and Vempala is known as Hit-and-Run \cite{Lovasz, hcorner}.}  This walk
mixes  in $O\left(n^3 (\frac{R}{r})^2 \ln \frac{R}{d\eps}\right)$
steps from a point at a distance $d$ from the boundary
\cite{hcorner}, where $\eps$ is the desired variation distance to
stationarity.

{ There has been previous work converting sampling algorithms into algorithms for optimization, notably the work of Bertsimas and Vempala \cite{BV, vemp_survey}.} 
Their algorithm can be be viewed as a randomized analogue of Vaidya's algorithm \cite{Vaidya}, wherein the role of the analytic center is played by the center of mass. In our work, we instead borrow the central idea of Karmarkar's algorithm, namely projective rescaling, to obtain a randomized algorithm for convex optimization. We need to perform this step just once, although affine rescaling occurs at every step.
 Generalizing prior work with Kannan on polytopes \cite{KNsample}, in Theorem~\ref{1tmain1} we present a Markov Chain for sampling from a convex body defined by logarithmic, hyperbolic and self-concordant barriers (see Section~\ref{secdefn}). When restricted to the case of polytopes, the bounds we present imply, up to universal constants, the bounds in  \cite{KNsample}.
In our case, the ``Dikin random walk" uses Gaussian steps having ellipsoidal covariance to make its individual steps, and is described in Section~\ref{s:walk}.

{ There is also work on sampling lattice points in polytopes based on sampling real points from the normalized Lebesgue measure in a polytope \cite{KV}.}

We believe that in cases where the solution is NP-hard, for example integer programming, randomized algorithms hold promise. While we are not aware of any advantage that our method possesses for the task of minimizing a linear form over a convex set over using an interior point algorithm associated with the barrier, we believe in the  utility of our randomized optimization scheme for integer programming. Huang and Mehrotra \cite{HuangMehrotra} have used a variety of geometric random walks 
for integer programming, including our random walk. The Dikin random walk (and a long-step variant of it) were implemented alongside the so-called ball walk and Hit-and-Run random walks by the authors of  \cite{HuangMehrotra}. In several cases our methods are competitive. In their approach instead of rounding the solution of an LP, they round the result of a biased random walk. Although deterministic algorithms perform better 
for solving LPs, it seems conceivable that the randomized algorithms produce better starting points for heuristics for solving integer programs. Another application of geometric random walks that we have not discussed is volume computation.
 The results of Huang and Mehrotra suggest that integer programming, or more generally, non-convex optimization are areas that could benefit from the use of geometric random walks.

In joint work with Alexander Rakhlin, we used the Dikin walk to obtain a {\it low regret} algorithm for online convex optimization (see \cite{NarRakh}). A property of the Dikin walk that was crucially used in that paper was that the bound on the conductance is "scale-free", when a metropolis rule is overlayed for sampling from exponential distributions.
This feature is not known to be shared by other random walks on convex sets, although the conductance of Hit-and-Run is known to be scale-free by the work of Lov\'{a}sz and Vempala \cite{hcorner} when sampling from the {\it uniform} distribution.  

{ A self-concordant barrier  $F$ (see \cite{Renegar}) } on a convex set $K$, is a convex function whose domain is the interior of $K$, that tends to infinity as one approaches its boundary, and whose second derivative at a point along any unit direction is large in a suitable sense compared to its first and third derivatives along the same vector. In order to convey the basic idea of Theorem~\ref{1tmain1}, let $D^2F(x)$ be the Hessian matrix of $F$ at $x$. We define the transition measure $P_x$ corresponding to $x$ to be roughly a Gaussian whose covariance matrix is a fixed multiple of $\left(D^2F(x)\right)^{-1}.$ The properties of the barrier  function cause the random walk to avoid the boundary, but at the same time cause it to take relatively large steps.
For example, let $K$ be the $2$-dimensional Euclidean ball $\{x : \|x\| \leq R\}$ and $F(x) := - \ln (R^2 - \|x\|^2)$ be a self-concordant barrier for it. Then, for $x \in K$, up to constants, the expected magnitude of the component of a step in the radial direction is ${R - \|x\|}$, while the expected magnitude of the component in the transverse direction is roughly $\sqrt{{R^2 - \|x\|^2}}$. It can be shown that the mixing time from  $0$ is independent of the diameter $2R$, and that the size and typical orientation of a step vary according to the local geometry.


In Theorem~\ref{1tmain2} we use this random walk to design an algorithm for optimization, which essentially consists of doing such a random walk on a projectively transformed version of $K$. This transformation  preferentially dilates regions corresponding to a larger objective value, causing them to occupy more space and hence become the target of a random walk.
In the case of polytopes, a slightly different version of this appeared in \cite{KNsample}. The Markov Chain considered in \cite{KNsample} was ergodic, while the one we use here is not. The analysis of the non-ergodic Markov Chain hinges upon the fact that it can be viewed as a limit of ergodic Markov Chains.
 Suppose $K$ is defined by a semidefinite constraint of rank at most $\nu_h$ and at most $m$ additional linear constraints, \ie $$K := \left\{\mathbf{x} \big | \mathbf{A}\mathbf{x} \leq \mathbf{b}\right\} \cap \{\mathbf{x} \big | \sum_i x_i \mathbf{A_i} \preceq \mathbf{B}\},$$ where $\mathbf{x} := (x_1, \dots, x_n)^T$, $\mA$ is an $m \times n$ matrix, and each $\mathbf{A_i}$ is a $\nu_h \times \nu_h$  symmetric positive definite matrix, as is $\mathbf{B}$. The symbol $\preceq$ indicates domination in the symmetric positive definite cone. Our results specialize to the following statement.

 Let $s \geq \frac{|p|}{|q|}$ for any chord $\overline{pq}$ of $K$ passing through a point $x \in K$. Let $\rho_t$ stand for the
 density of the $t^{th}$ point $x_t$ of the Dikin walk. Then, after $$t = O\left(n (m + n \nu_h) \left(n  \ln((m + n \nu_h) s) +  \ln \frac{1}{\eps}\right)\right)$$  steps are taken by a Dikin walk starting at $x$, the total variation distance and the $\mathcal{L}_2$ distance of the density $\rho_t(\cdot)$ of the point  to the uniform density are less than $\eps$.

One technical contribution of this paper is a family of lower bounds (see Theorem~\ref{1thmche}) for the  isoperimetric constants of  weighted Riemannian manifolds on which interior point methods perform a kind of steepest descent.
While the results in \cite{KNsample} used the isoperimetric properties of the ``Hilbert metric," we use a Riemannian metric. It is not clear whether the Hilbert metric can be used to obtain the results in this paper,  because we consider convex sets defined by a combination of three barriers of increasingly general type - logarithmic, hyperbolic and self-concordant. These barriers are defined in Section~\ref{secdefn}. The logarithmic, hyperbolic and self-concordant barriers give rise to  metrics possessing progressively weaker regularity properties. The net metric that we are (seemingly) forced to use is a weighted sum of the three metrics, where the weights take care of this progressively weakening regularity. It is unclear how the somewhat disparate issues that arise from the use of a combination of these different kinds of metrics could be dealt with in a unified way if one directly used the Hilbert metric.
The proof of our mixing bounds follows the broad outline of most results on mixing in geometric random walks. We first prove a purely geometric isoperimetric inequality for a ``Dikin manifold". We then obtain a bound on the probabilistic distance between the one-step densities starting from two nearby points.

We end this section with an outline of the paper.
Section~\ref{sec:relwork} describes past work on sampling convex sets.
Section~\ref{secdefn} describes self-concordant barriers, which are needed to define the Dikin walk.
Section~\ref{sec:oracle} defines the oracle model that will form the framework of our algorithm.
Section~\ref{sec:main} describes the three main theorems, Theorem~\ref{1tmain1}, Theorem~\ref{thm:bob} and Theorem~\ref{1tmain2}. Section~\ref{sec:metric} introduces a ``Hessian metric" based on a self-concordant barrier. Section~\ref{sec:analysis} describes the proof of Theorem~\ref{1tmain1}. Section~\ref{sec:direct} describes the proof of Theorem~\ref{thm:bob}. Section~\ref{sec:convex} describes the proof of Theorem~\ref{1tmain2}. We have provided the proofs of lemmas in the Appendix.


\section{Comparison with related work}\label{sec:relwork}
We begin with some definitions that are used in related work.

\begin{definition}[Warm start] Let $\rho_0$ be a starting distribution for a Markov Chain $x_0, x_1, \dots$ whose stationary distribution is the uniform distribution $\mu$ on a convex set $K$. We say $\rho_0$ is a warm start for the chain if there is a fixed universal constant $C$ such that
$$\sup_A \frac{\rho_0(A)}{\mu(A)} < C,$$ where the supremum is taken over all measurable subsets $A$.
\end{definition}
\begin{definition}[Mixing time]Let $\rho_0$ be a starting distribution for a Markov Chain $x_0, x_1, \dots$ whose stationary distribution is the uniform distribution $\mu$ on a convex set $K$. Let $\rho_t$ be the distribution of the $t^{th}$ point of the Dikin walk.  We define the mixing time in total variation distance $T_{TV}(\eps, \rho_0)$ to be the smallest $t$ such that
the total variation distance of $\rho_t(\cdot)$ to the uniform distribution on $K$ is less than $\eps$.
 
  For every bounded $f$, let $\|f\|_{2, \mu}$ denote
$\sqrt{\int_K f(x)^2 d\mu(x)}$.
Let $f_t(x):= \rho_t(x) - (\vol(K))^{-1}.$
We define the mixing time in $\LL_2$ distance $T_{2}(\eps, \rho(x_0))$ to be the smallest $t$ such that
 $$\|f_t\|_{2, \mu} < \eps.$$
\end{definition}

 Let $B(x, \tau)$ be defined to be the $n-$dimensional Euclidean ball of radius $\tau$ centered at $x$ and suppose $K$ is an $n$-dimensional convex set such that $B(0, r) \subseteq K \subseteq B(0, R)$.  The Markov Chain known as the  ``Ball walk" \cite{LS1}, \cite{KLS} is defined as follows. If the random walker is at a point $x_i$ in a convex body $K$ at time step $i$, a random point $z$ is picked in $B(x_i, O(\frac{1}{\sqrt{n}}))$, and  $x_{i+1}$ is set to $z$ if it lies in $K$, otherwise the move is rejected and $x_{i+1}$ is set to $x_i$.
{The mixing time of this walk from a warm start in order to achieve a constant total variation distance to stationarity is $O^*\left( \frac{n^2R^2}{r^2}\right)$ (where $O^*(T)$ is synonymous with $T \log^{O(1)} T$).
More recently, a random walk known as Hit-and-Run, was analyzed in \cite{Lovasz} and \cite{hcorner}.} If the random walker is at a point $x_i$ in a convex body $K$ at time step $i$, a vector is picked from the uniform distribution on the sphere and through $x_i$, and $x_{i+1}$ is chosen from the uniform measure on the chord $\{x_i + \l v \big| \l \in \RR\} \cap K$.  Unlike the Ball walk, this walk provably mixes rapidly from any interior point, with a weak (logarithmic) dependence on the distance of the starting point from the boundary.
From a warm start, the mixing time of Hit-and-Run  is $O\left(\frac{n^2R^2}{r^2}\right)$, and its mixing time from a fixed point at a distance $d$ from the boundary is $O\left(n^3 (\frac{R}{r})^2 \ln \frac{R}{d}\right)$.

The mixing time of the Dikin walk in two cases of interest are as follows (details are provided in Theorem~\ref{1tmain1}). Let $x \in K$ and for all chords $\overline{pq}$ passing through  $x$, $\frac{|p-x|}{|q-x|} \leq s$. We will call such a point $x$, $s-$central.
 Suppose $K \in \RR^n$ is
\ben \item   the intersection of an $n-$dimensional affine subspace (identified with $\RR^n$) with the semidefinite cone $S^{\nu_h \times \nu_h}$ of $\nu_h \times \nu_h$ matrices endowed with the hyperbolic barrier $F(x) = - \ln \det x$ or
\item the intersection of $m = \frac{\nu_h}{2}$ ellipsoids, $A_1 B \cap A_2B \cap \dots \cap A_{m} B$ where $A_i$ are non-singular affine transformations and $B$ is the Euclidean Ball. In this case, the hyperbolic barrier is $F(x) = - \sum \ln( 1 - \|A_i^{-1}(x)\|^2).$
\een
Then we define the complexity parameter of the barriers as $\nu = n \nu_h$ and the mixing time starting at $x$ is $O( n \nu(n  \ln(\nu s)+ \ln \frac{1}{\eps}))$.
Whether or not the bodies are in isotropic position, in the above cases 1. and 2. corresponding to semidefinite and quadratic programs, when $\nu_h = O(n^{1-\eps})$ { [and so $\nu = O(n^{2- \eps})$ and the mixing time from a warm start is $O(n^{3-\eps})$],} the mixing time bounds are an improvement over  the existing  bound for Hit-and-Run \cite{hcorner}, which is $O(n^3)$ (assuming $R/r = \Omega (\sqrt{n})$) .

Lov\'{a}sz \cite{Lovasz} proved a lower bound of $\Omega(n^2 p^2)$ on the mixing time of Hit-and-Run in a cylinder $B_n \times [-p, p]$ from a warm start, where $B_n$ is the unit ball in $n-$dimensions.
Dikin walk has a mixing time of $O(n^2)$ from a warm start.  
Thus for a cylinder with $p = \omega(1)$, the lower bound on the number of steps needed for  Hit-and-Run to mix (without rescaling the body) is larger than the upper bound on the number of steps for Dikin walk.




{ In the specific case where the constraints are either semidefinite or linear, we can compare the upper bounds on the number of arithmetic operations needed for one step in Hit-and-Run and Dikin walk. This has been done in Section~\ref{sec:lin_sem}.}

\section{Self-concordant barriers}\lab{secdefn}
 Let $K$
be a convex subset of $\RR^n$ that is not contained in any
$n-1$-dimensional affine subspace and $int(K)$ denote its interior.
For any function $F$ on $int(K)$ having continuous derivatives of order $k$,
for vectors $h_1, \dots, h_k \in \RR^n$ and $x \in int(K)$, for $k \geq 1$, we recursively define
$$D^kF(x)[h_1, \dots, h_k]  := $$ $$\lim_{\eps \ra 0 } \frac{D^{k-1}F (x + \eps h_k) [h_1, \dots, h_{k-1}] - D^{k-1}F (x) [h_1, \dots, h_{k-1}]}{\eps},$$
where $D^0F(x) := F(x)$.
Following Nesterov and Nemirovskii \cite{NNbook}, we call a real-valued function $F:int(K) \ra \RR$,  a
regular self-concordant barrier if it satisfies the
conditions stated below. For convenience, if $x \not \in int(K)$, we define $F(x) = \infty$.
\begin{enumerate}
\item (Convex, Smooth) $F$ is a convex thrice continuously differentiable function on $int(K)$.
\item (Barrier) For every sequence
of points $\{x_i\} \in int(K)$ converging to a point $x \not \in
int(K)$,  $\lim_{i \ra \infty} f(x_i) = \infty$.
\item (Differential Inequalities)    For all $h \in \RR^n$ and all $x \in int(K)$, the following
inequalities hold.

\begin{enumerate}

\item $D^2 F(x)[h, h]$ is $2$-Lipschitz continuous with respect to the local
norm, which is equivalent to $$D^3F(x)[h, h, h] \leq 2 (D^2F(x)[h, h])^{\frac{3}{2}}.$$ \een
\item $F(x)$ is $\nu$-Lipschitz continuous with respect to the local norm
defined by $F$,
$$|DF(x)[h]|^2 \leq \nu D^2F(x)[h, h].$$ We call the smallest non-negative real number $\nu$ for which this holds for all $h \in \RR^n$ and $x$ in the interior of $K$, the self-concordance
parameter of the barrier.
 \end{enumerate}
 It follows from these conditions that if $F$ is a self-concordant barrier for $K$ and $A$ is a non-singular affine transformation, then $F_A(x) :=F(A^{-1}x)$ is a self-concordant barrier for $AK$.  This fact is responsible for the affine-invariance of Dikin walk.
 Some  examples of convex sets for which explicit barriers are known are
 \ben
\item Convex sets defined by hyperbolic constraints. This set includes sections of semidefinite cones. Polytopes and the intersections of ellipsoids can be expressed as sections of semidefinite cones.
\item Convex sets defined by the epigraphs of matrix norms (see page 199 of \cite{NNbook}).
\een
For other examples and methods of constructing  barriers for new convex sets by combining existing barriers, see Chapter 5 of \cite{NNbook}.
\subsection{Generic self-concordant barrier} We refer by $F_s$ to a generic self-concordant barrier.
 \subsection{Hyperbolic barriers}

  \begin{definition} A homogenous polynomial $p:\mathbb{C}^n \ra \mathbb{C}$ is called hyperbolic with respect to a direction $d \in \mathbb{R}^n$ if $p(d) \neq 0$  and there exists a constant $t_0 \in \R$ such that $p(x + itd) \neq 0$ if $x \in \R^n$ and $t < t_0$.
  \end{definition}
  Associated with such a polynomial is a cone of hyperbolicity (see \cite{guler}).
  The function $$F_h(x) = -\log p(x)$$ is called a hyperbolic barrier.

  For the concrete applications in this paper, it suffices to note that on the semidefinite cone $S^{m \times m}$, $-\ln \det x$ is a hyperbolic barrier with parameter $m$, and that on the intersection of  ellipsoids, $A_1 B \cap A_2B \cap \dots \cap A_{m} B$ where $A_i$ are non-singular affine transformations and $B$ is the Euclidean Ball, $ - \sum \ln( 1 - \|A_i^{-1}(x)\|^2)$ is a hyperbolic barrier with self-concordance parameter $m$.

 \begin{lemma}[Theorem 4.2, G\"{u}ler \cite{guler}]\label{lem:1} If $F$ is a hyperbolic barrier,
 $$|D^4 F(x) [h, h, h, h]| \leq 6 \left(D^2 F(x)[h, h]\right)^2.$$
 \end{lemma}

\subsection{Logarithmic barrier of a polytope}
Given any set of linear constraints $\{a_i^T x \leq 1\}_{i = 1}^m$, the logarithmic barrier is a real valued function defined on the intersection of the halfspaces defined by these constraints, and is given by $$F_\ell(x) = - \sum_{i=1}^m \ln\left(1 - a_i^T x\right).$$

\subsection{Dikin Ellipsoids}

Around point $x \in K$, given a self-concordant barrier $F$ the {\it Dikin ellipsoid} of radius $r$ is defined to be
 $$D_x := \{ y : D^2 F(x)[x - y, x - y] \leq r^2\}.$$
 Dikin ellipsoids are affine
invariants in that, if the Dikin ellipsoid of radius $r$ around a point $x \in K$ is $D^r_x$ and $T$ is a non-singular affine transformation of $K$,
the Dikin ellipsoid of radius $r$ centered at the point $Tx$ for
$T(K)$ is $T(D^r_x)$, as long as the new barrier that is used is $H(y) := F(T^{-1} y)$.

The following fact will be used in the proof of  Lemma~\ref{l:h} and Lemma~\ref{l:s}.
\begin{fact}\lab{f:s1}
For any $y$ such that $$D^2F(x)[x - y, x - y] = r^2 < 1,$$
 for any vector $h \in \RR^n$,
 \beqs (1 - r)^2 D^2F(x)[h, h] \leq D^2F(y)[h, h] \leq \frac{1}{(1 - r)^2}D^2F(x)[h, h]. \eeqs
  Also, the Dikin ellipsoid centered at $x$, having
radius $1$, is contained in $K$. This has been shown in Theorem
2.1.1 of Nesterov and Nemirovskii \cite{NNbook}.
\end{fact}
The following was proved in a more general context by Nesterov and
Todd in Theorem 4.1, \cite{NT2}.
\begin{fact}[Nesterov-Todd]\label{thm:NT2}
Let $\overline{pq}$ be a chord of a polytope $P$ and $x, y$ be interior points
on it so that $p, x, y, q$ are in order. Let $D_x$ be the Dikin ellipsoid of unit radius at $x$ with respect to a point $x$. Then $z \in D_y$ implies
that $p + \frac{|p-x|}{|p-y|}(z - p) \in D_x.$
\end{fact}

\section{Oracle model}\label{sec:oracle}
There are two  standard information models for convex sets in the operations research literature, the separation model and the (self-concordant) barrier model (see Freund \cite{freund}, page 2). Existing work on sampling convex sets, with the exception of Kannan and Narayanan \cite{KNsample} has focussed on the separation model and a weaker model known as the membership oracle model. The self-concordant barrier model we will consider is the following.
\begin{enumerate}
\item We are guaranteed that the origin belongs to $K$ and that $K$ has a self-concordant barrier $F$ (see Section~\ref{secdefn}).

    { \item   We are given a real number $s$ such that for any chord $\overline{pq}$ of $K$ such that $\frac{|p|}{|q|} \geq 1$ through the origin,  $ \frac{|p|}{|q|} \leq s$ (see Figure 1).}
\item On querying a point $x \in \RR^n$, we are returned a positive semidefinite matrix corresponding to the Hessian of $F$ at $x$, if $x \in K$ and returned ``No" if $x \not \in K$.
    \end{enumerate}

\subsection{Presentation of the convex set $K$}\lab{ss:pres}
For the definitions of logarithmic, hyperbolic and self-concordant barriers, we refer to Section~\ref{secdefn}.
We will assume that the convex set $K$ is specified as the set of points that satisfy a family of constraints
$$K := \bigcap_{i=1}^m \{F_i(x) < \infty\}$$ where the $F_i$ are either logarithmic, hyperbolic or arbitrary self-concordant functions. Without loss of generality, we may aggregate these barriers and may assume that
 $K := K_\ell \cap K_h \cap K_s$, where $K_\ell$ is a polytope with $m$ faces accompanied by the logarithmic barrier ${F_\ell}$, $K_h$ is a convex set accompanied by a hyperbolic barrier ${{F_h}}$ with self-concordance parameter $\nu_h$, and $K_s$ is a convex set accompanied by a self-concordant barrier ${{F_s}}$ whose self-concordance parameter is $\nu_s$. Although their intersection is bounded, each of these convex sets may be unbounded.
 \begin{definition}
  Given $F_\ell, F_h, F_s$ as above, we define  the self-concordant barrier function  $$F := {F_\ell} + n  {F_h} + n^2  {F_s}, $$ and define \beq\lab{eq:compl}\nu := m + n  \nu_h  + (n \nu_s)^2\eeq to be the {\it complexity parameter} of $F$ (which is different from its {\it self-concordance parameter} the latter being bounded above by $m + n\nu_h + n^2 \nu_s$ ).
 \end{definition}

  Let $C$ be a sufficiently large universal constant. We define the radius of a Dikin step, $r$ to be $1/C$.
 For a point $x \in K$ and $v \in \RR^n$, we define
 $$\|v\|_x^2 := D^2 F(x)[v, v].$$
  The random walk we use here is a variation  of the Dikin walk defined in \cite{KNsample}. Instead of picking the next point from a Dikin ellipsoid, here we pick it from a Gaussian having that covariance.

\section{Main Results}\label{sec:main}
In this section we present our main results.
Theorem~\ref{1tmain1} shows that given a convex body accompanied with a barrier, it is possible to use the Hessian of the barrier to define a rapidly mixing random walk. This allows us to efficiently pick a nearly random point from the uniform measure on the convex set.

Theorem~\ref{thm:bob} shows that in a special case where the convex set is a direct product of convex sets,
the mixing time is governed by the worst factor. While we are not aware of any algorithmic application of this result, since it is easy to sample from a direct product, given samples from the factors, this illustrates that one can do better than to apply the Localization Lemma (\cite{LS1}) in a natural special case.

The mixing results can be adapted to give a random walk based polynomial-time Las Vegas algorithm for optimizing a linear function $c^T x$ on certain convex sets $K$. The complexity of this algorithm is roughly the same as that of the sampling algorithm. This is the content of Theorem~\ref{1tmain2}.

\subsection{Sampling through Dikin Walk}\lab{s:walk}

 For $x \in int(K)$, let $G_{x}$ denote the Gaussian density function given by

$$G_{x}(y) := \left(\frac{n}{ \pi r^2}\right)^{\frac{n}{2}}  \exp \left(-\frac{n\|x - y \|_x^2}{r^2} + V(x)\right),$$ where
$$V(x) = \left(\frac{1}{2}\right) \ln  \det  D^2F(x).$$

Let $r$ be a sufficiently small absolute constant {(asymptotically less or equal to $\frac{1}{\sqrt{2}}$)} so that the following tail estimate for the Gaussian holds.

\begin{fact} Suppose $x \ra z$ is a transition of the Dikin walk given by
$$G_{x}(z) := \left(\frac{n}{ \pi r^2}\right)^{\frac{n}{2}}  \exp \left(-\frac{n\|x - z \|_x^2}{r^2} + V(x)\right),$$
then, $$\p\left[\|x - z\|_x \leq \frac{1}{2}\right] \geq 1 - 10^{-3}.$$
\end{fact}
\begin{figure}\lab{fig:set}
\begin{center}
\includegraphics[height=2.0in]{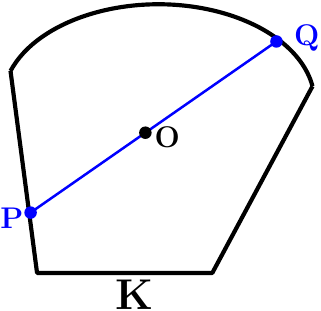}
\end{center}
\caption{A chord $\overline{pq}$ in a convex body $K$}
\end{figure}

The Dikin Walk Algorithm is given below.
\begin{algorithm}{Dikin Walk Algorithm}\\
\noindent Let $0 = x_0 \in int(K)$. For $i \geq 1$, given $x_{i-1}$,
\begin{enumerate}
\item Toss a fair coin. If \verb"Heads" let $x_i := x_{i-1}$.
\item Else \ben
\item Choose $z$ from the density $G_{x_{i-1}}$.
\item If $ z \in K$, let $$x_i := \left\{
                    \begin{array}{ll}
                      z, & \hbox{  \text{with probability} $\min\left(1,\,\, \frac{G_{z}(x_{i-1})}{G_{x_{i-1}}(z)} \right)$} \\ 
                      x_{i-1} & \hbox{otherwise.}
                    \end{array}
                  \right.$$
\item If $ z \not\in K$, let $x_i := x_{i-1}$.
\een
\end{enumerate}
\end{algorithm}

\begin{figure}\label{1fig:set}

\end{figure}
We prove the following in Subsection~\ref{ssec:1tmain1}.
\begin{theorem}[Sampling]\label{1tmain1}
 Let $K \ni 0$ be an $n-$dimensional convex set accompanied by a barrier $F$ as in Subsection~\ref{ss:pres}, with complexity parameter $\nu$. Let $s \geq \frac{|p|}{|q|}$ for any chord $\overline{pq}$ of $K$ containing the origin, with $\frac{|p|}{|q|}\geq 1$. Let $e$ be the time of the first non-trivial move of the Markov chain. Then,  the number of  steps $x_t$ after $e$, before both the distances --  total variation distance and the $\mathcal{L}_2$ distance of the density $\rho_t(x)$ of $x_t$ to the uniform density -- are less than $\eps$ is {$O\left(n \nu \left(n  \ln( \nu s) +  \ln \frac{1}{\eps}\right)\right).$ }The number of steps needed from a warm start is
$O\left(n \nu  \ln \frac{1}{\eps}\right).$ The time of the first non-trivial move $e$ has a geometric distribution whose mean is bounded above by a universal constant.
 \end{theorem}

In particular, suppose $K$ is
\ben \item[(S)] a slice of the semidefinite cone $S^{\nu_h \times \nu_h}$ of $\nu_h \times \nu_h$ matrices with $F(x) = - \ln \det x$ or
\item[(Q)] the intersection of $\frac{\nu_h}{2}$ ellipsoids, $A_1 B \cap A_2B \cap \dots \cap A_{\nu_h/2} B$ where $A_i$ are non-singular affine transformations and $B$ is the Euclidean Ball. In this case, $F(x) = - \sum \ln( 1 - \|A_i^{-1}(x)\|^2)$.
\een
In each case the complexity parameter $\nu$ is $n\nu_h$ and
the mixing time from a fixed ``$s-$central" point or a warm start, respectively, are
$O\left(n \nu\left(n  \ln( \nu s) +  \ln \frac{1}{\eps}\right)\right)$
and
$O\left(n \nu  \ln \frac{1}{\eps}\right).$

The mixing bounds in this paper are obtained by relating the Markov Chain to the metric of  a Riemannian manifold studied in operations research \cite{NN} and \cite{NT}, rather than the Hilbert metric \cite{KNsample} and \cite{Lovasz}. The aforementioned Riemannian metric possesses several potentially useful characteristics. For example, when the convex set is a direct product of convex sets, this metric factors in a natural way into a product of the metrics corresponding to the individual convex sets, which is not the case for the Hilbert metric.   Using results of Barthe \cite{barthe} and Bobkov and Houdr\'{e} \cite{bobkov} on the isoperimetry on product manifolds, this leads to an improved upper bound on the mixing time when $K$ is a  direct product of convex sets, and opens up the future possibility of using differential-geometric techniques for proving isoperimetric bounds, in addition to relying  on the Localization Lemma, which underlies the analysis of  all Markov Chains on convex sets ever since it was introduced in (Lov\'{a}sz and Simonovits \cite{LS1}). Even if $K$ is a direct product of convex sets, the  Dikin Markov Chain itself does not factor into a product of Dikin Markov Chains
and  Theorem~\ref{thm:bob} does not follow from a direct use of the Localization Lemma.
\begin{theorem}\lab{thm:bob}
If an $n-$dimensional convex set $K := K_1 \times \dots \times K_h$ is the direct product of convex sets $K_i$, each of which  individually has a function $F_i$ with a complexity parameter (defined in Equation~\ref{eq:compl}) at most $\nu$, then,  the mixing time of Dikin walk from a warm start on $K$ defined using the function $\sum_{i=1}^h F_i$ is $O(\nu n)$.
\end{theorem}
 When there are $h = \Omega(n)$ factors, each of which is a polytope with $\kappa$ faces,  the total number of faces of $K$ is $\nu_\ell = \Omega(n\kappa)$. In this case,   the results of  \cite{KNsample} give a bound of $O(n \nu_\ell)$ while Theorem~\ref{thm:bob} gives a bound of $O(\nu_\ell)$.
\begin{figure}\label{1fig:walk}
\begin{center}
\includegraphics[height=2.0in]{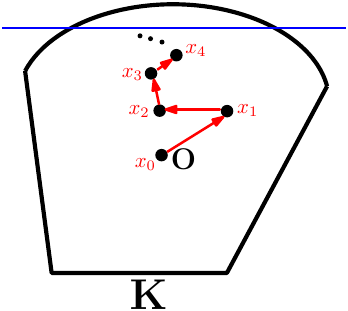}
\end{center}
\caption{Trajectory of a Dikin walk for optimizing a linear function on a convex set}
\end{figure}

\subsection{Convex programming}\lab{sec:convex_prog}


 Our algorithm for convex optimization in this paper is a Las Vegas algorithm rather than a Monte Carlo algorithm (as was the case in \cite{KNsample}). It is also different from \cite{KNsample} in that  the Markov Chain used here does not depend on $\eps$, the error tolerance.

We will consider convex programs specified as follows.
Suppose we are given a convex set $K$ containing the origin as an interior point and a linear objective $c$, such that
  $$Q := K \cap \{y : c^Ty \leq
1\}$$ is bounded, for any chord $\overline{pq}$ of $Q$ passing through the origin, $\frac{|p|}{|q|} \leq s$    and $\eps
> 0$ (if $B(0, r) \subseteq K \subseteq B(0, R)$, then $s \leq \frac{R}{r}$). Then, the algorithm is required to do the following.
If $\exists \, x \in K$ such that $c^T x \geq 1$,
  output $x' \in K$ such that $c^T x' \geq 1 -\eps$.

Let $T: Q \ra \RR^n$ be defined by
$$T(x) = \frac{ x}{ 1 - c^T x},$$ and let $\tf$ be a barrier for $\tk := T(Q)$. Such a barrier can be easily constructed from $F$; details appear in Section~\ref{ssec:const_bar}.
\subsubsection{Constructing barriers}\lab{ssec:const_bar}
Let $T: Q \ra \RR^n$ be defined by
$T(x) = \frac{ x}{ 1 - c^T x}$ and $\tk$ be the projective image $TQ$ of $Q$ as {defined in} Section~\ref{sec:convex_prog}. The construction in \cite{Nest_selfconc}, provides us with a barrier $\tf_s$ on $\tk$.  The barrier $\tf_s$ is given by
    $$\tf_s(y) := \left(\frac{8}{3 \sqrt{3}} + \frac{1}{2\sqrt{\nu_s}} \left( \frac{7}{3}\right)^{\frac{3}{2}}\right)^2 \left({F_s}\left(\frac{y}{1 + c^T y}\right) + 2 \nu_s \ln(1 + c^Ty)\right),$$ whose self-concordance parameter is $\leq (3.08 \sqrt{\nu} + 3.57)^2$.
If ${F_h} = - \ln p(x)$ where $p$ is a hyperbolic polynomial of degree $\nu_h$, $\tf_h$ is defined simply by
$$\tf_h(y) := {F_h}\left(\frac{y}{1 + c^T y}\right) +  \nu_h \ln(1 + c^Ty),$$ and has the same self-concordance  parameter $\nu_h$.
This applies to the special case of the { logarithmic barrier as well.}
%
For any point $x \in int(\tk)$, we use the Hessian matrix $D^2\tf$ to define a norm $$\hat{\|v\|}_x := (v^T D^2\tf v)^{\frac{1}{2}}.$$

{ $$\tg_{x}(y) := \left(\frac{n}{ \pi r^2}\right)^{\frac{n}{2}}  \exp \left(-\frac{n\widehat{\|x - y \|}_x^2}{r^2} + \tv(x)\right),$$} where
$$\tv(x) = \left(\frac{1}{2}\right)\ln  \det  D^2\tf(x).$$
For $x \not\in int(\tk)$, for any $y \neq x$, $\tg_x(y) := 0 $.
Let $$s := \sup_{\overline{pq} \ni O} \frac{|p|}{|q|},$$ where the supremum is taken over all chords of $K$ containing the origin.


Our algorithm for convex optimization consists simply of doing a modified Dikin walk {(see Figure 2)} on $\tk$ for a sufficient number of steps that depends on the desired accuracy $\eps$ and confidence $ 1 - \de$.
Note that we define $\tg_t$ using $\tf$ in the same way that $G_t$ was defined using $F$.

\subsubsection{Las Vegas Algorithm}
We state the Las Vegas Algorithm below.
\begin{algorithm}{Las Vegas Algorithm for Optimization}\\
Let $x_0 = 0 $. While  $c^T x_{i-1} < 1 - \eps$,
\ben
\item Toss a fair coin. If \verb"Heads", set $x_i := x_{i-1}$.
\item Else,
\begin{enumerate}
\item Choose $z$ from the density $\tg_{T(x_{i-1})}$.
\item If $ z \in \tk$, let $$x_i := \left\{
                    \begin{array}{ll}
                      T^{-1}(z), & \hbox{  \text{with probability} $\min\left(1,\,\, \frac{\tg_{z}(T(x_{i-1}))}{\tg_{T(x_{i-1})}(z)} \right)$} \\ 
                      x_{i-1} & \hbox{otherwise.}
                    \end{array}
                  \right.$$
\item If $ z \not\in \tk$, let $x_i := x_{i-1}$.
\end{enumerate}
\een
\end{algorithm}

\begin{theorem}\label{1tmain2}
 Let $K, F$ and $s$  be as in Theorem~\ref{1tmain1}.  In the cases where $F$ is a $\nu$-barrier or a hyperbolic barrier with parameter $\nu$, let $\tau(\eps, \de)$ be set to { $O\left(n \nu \left( \ln \frac{1}{\de} + \left(n \ln \frac{s n \nu}{\eps}\right)\right)\right).$}  If $\{c^T x \geq 1\} \cap K$ is nonempty and $x_0, x_1, \dots$ is the modified Dikin walk in Las Vegas Algorithm, then \beqs\p\left[\forall i > \tau(\eps, \de),\, c^T x_i  \geq 1 - \eps \right] \geq 1 - \de.\eeqs
\end{theorem}

We remind the reader that for a Las Vegas algorithm, correctness is guaranteed, but there are no absolute bounds on the run time. Consider the stopping rule under which the random walk is terminated the first time that it reaches  $c^T x_i  \geq 1 - \eps$. The following corollary shows that the resulting algorithm is a Las Vegas algorithm.
\begin{corollary}[Las Vegas algorithm for optimization] For any $\kappa > 0$, with probability $1$,
$$\lim_{i \ra \infty} e^{ i^{1 -\kappa}} (1 - c^T x_i) = 0.$$
\end{corollary}

\section{Isoperimetry}
\subsection{Metric defined by a  barrier}\label{sec:metric}
The isoperimetric properties of a certain metric measure space govern the mixing bounds in the present work.
In this section, we provide preliminary definitions and properties needed in our proofs.

For any smooth strictly convex function $G$, the Hessian $D^2 G$
is positive definite.
\begin{definition} Given the barrier $F$, for every $x \in supp(F)$  and $u, v \in \RR^n$, $ D^2F(x)[u, v]$ is bilinear, and  $\sqrt{<_Fu, u>_x}$ is a norm.  We define
\ben \item $<_F u, v>_x := D^2{F}(x)[u, v]$ and $\|_F u\|_x := \sqrt{<_F u, u>_x}$,
\item $<_\ell  u, v>_x := D^2{F_\ell}(x)[u, v]$ and $\|_\ell u\|_x := \sqrt{<_\ell  u, u>_x}$,
\item $<_h u, v>_x := D^2{F_h}(x)[u, v]$ and $\|_h u\|_x := \sqrt{<_h u, u>_x}$ and
\item  $<_s u, v>_x := D^2{F_s}(x)[u, v]$ and $\|_s u\|_x := \sqrt{<_s u, u>_x}$.
  \een
\end{definition}
 \begin{definition}
 We define $$d_F(x, y) = \inf_\Gamma \int_z \|_Fd \Gamma\|_z dz $$ where the infimum is taken over all rectifiable paths $\Gamma$ from $x$ to $y$. $d_\ell, d_h$ and $d_s$ are defined analogously in terms of the respective norms $\|_\ell\cdot\|$, $\|_h \cdot \|$ and $\|_s\cdot\|$, which we refer to as Dikin norms.
 \ben
 \item $d_\ell(x, y) = \inf_\Gamma \int_z \|_\ell d \Gamma\|_z dz.$
 \item $d_h(x, y) = \inf_\Gamma \int_z \|_h d \Gamma\|_z dz.$
 \item $d_s(x, y) = \inf_\Gamma \int_z \|_s d \Gamma\|_z dz.$
 \een
 Where $F$ is clear from context, the subscript $F$ will be skipped.
 \end{definition}

 Let $\MM$ be the metric space associated with $F$ whose point set is $K$ and metric is $d_F$.

The following lemma is needed to relate the Riemannian metric $d_s(x, y)$ to the Dikin norm $\|_sx - y\|$.
\begin{lemma}[Nesterov-Todd (Lemma 3.1 \cite{NT})]\label{1lNT} If $\|_sx - y\|_x < 1$ then, $$\|_sx - y\|_x - \|_sx - y\|_x^2 \leq d_s(x, y) \leq - \ln (1 -  \|_sx -y\|_x).$$ \end{lemma}
Since any logarithmic or hyperbolic barrier is also a self-concordant barrier, this implies that
if ${\|_F}x - y\|_x < 1$ then, $${\|_F}x - y\|_x - {\|_F}x - y\|_x^2 \leq d_F(x, y) \leq - \ln (1 -  {\|_F}x -y\|_x).$$
While some of the presented  bounds  can be obtained from the isoperimetric bounds for the ``Hilbert metric" (Theorem~\ref{1thm:l}) proved by Lov\'{a}sz, we can prove stronger results for sampling certain classes of convex sets such as the direct product of an arbitrary number of convex sets, by using results of Barthe \cite{barthe} and Bobkov and Houdr\'{e} \cite{bobkov} on the isoperimetry of product spaces, which do not seem to follow directly from the Hilbert metric. In particular, for a direct product of an arbitrary number of polytopes, each defined by $O(\kappa)$ constraints,  this allows us to show an upper bound on the  mixing time from a warm start of $O(\kappa n)$. The bound obtained using the Hilbert metric in the obvious way is $O(\kappa n^2)$, since the Hilbert metric on a direct product does not decompose conveniently into factors as does the Riemannian metric.

Riemannian metrics defined in this way have been studied because of their
importance in convex optimization, for example, by Nesterov and Todd in
\cite{NT} and by Nesterov and Nemirovski in \cite{NN}, and Karmarkar studied the properties of a related metric \cite{Kar2} that underlay his celebrated algorithm \cite{Kar1}. For other work on sampling Riemannian manifolds motivated by statistical applications, { see  \cite{lebeau} and  \cite{NarNiy}, and  Chapter 8 of \cite{thesis}} and the references therein.
\subsection{Results on Isoperimetry}
Let $\MM$ be a metric space endowed with distance function $d$ and $\mu$ be a probability
measure on it. We term $(S_1, \MM \setminus (S_1 \cup S_2),
S_2)$ a $\delta$-partition of $\MM$, if $$\de \leq d(S_1, S_2) := \inf\limits_{x \in S_1, y \in S_2} d(x, y),$$ where $S_1$, $S_2$ are
measurable subsets of $\MM$. Let $\PP_\delta$ be the set of all
$\delta$-partitions of $\MM$. The
isoperimetric constant $\C(\de, \MM, \mu)$ is defined as
$$\inf_{
\PP_\delta} \frac{\mu(\MM \setminus (S_1 \cup
S_2))}{\mu(S_1) \mu(S_2)}.$$

Given interior points $x,y$ in  $int(K)$, suppose $p,q$ are the
ends of the chord in $K$ containing $x,y$ and $p,x,y,q$ lie in that
order.
\begin{definition}Denote by $d_H$ the Hilbert (projective) metric defined by $$d_H(x, y) := \ln\left(1 + \frac{|x-y||p-q|}{|p-x||q-y|}\right).$$   

\end{definition}

\begin{definition}
For $x \in K$ and a vector $v$, $|v|_x$ is defined to be $$\sup_\alpha \{ x \pm \alpha v \in K\}.$$
\end{definition}

Let $\C := \C(\de, \MM, \mu),$ where $\de = \frac{1}{\sqrt{n}}.$

\begin{theorem}[Theorem 2.3.2 (iii), \cite{NNbook}]\label{1tself}Let $F$ be a  self-concordant barrier whose self-concordance parameter is $\nu_s$ as defined in Section~\ref{secdefn}. Then,
for all $h \in \RR^n$ and $x \in int(K)$
$$|h|_x \leq \|_sh\|_x \leq 2(1 + 3 \nu_s)|h|_x.$$
\end{theorem}
The following result is implicit in \cite{guler}.
\begin{theorem}[G\"{u}ler, \cite{guler}]\label{1thyp}
Let $ - \ln p(x)$ be a hyperbolic barrier  for $K$, where $p$ has degree $\nu_h$.
Then,  for all $h \in \RR^n$ and $x \in int(K)$, $$|h|_x \leq \sqrt{D^2 {F_h} (x) [h, h]} \leq \sqrt{\nu_h} |h|_x.$$
\end{theorem}
\begin{lemma}\lab{l:1n} \ben \item ${d_{s}(x, y)} \leq 2(1 + 3\nu_s)  d_H(x, y)$.
\item
   $d_{h}(x, y) \leq \sqrt{ \nu_h} {d_H(x, y)} $.
   \item ${d_{\ell}}(x, y)\leq \sqrt{m} {d_H(x, y)} $.
   \item $d_F(x, y) \leq O(\sqrt{\nu}) d_H(x, y)$.
   \een
    \end{lemma}
\begin{proof}
For any $z$ on the segment $\overline{xy}$, $d_H(x, z) + d_H(z, y) = d_H(x, y)$. Therefore it suffices to prove the result infinitesimally.  By Lemma~\ref{1lNT} $$\lim_{y \ra x}\frac{d_F(x, y)}{\|_Fx - y\|_x} = 1,$$ and
a direct computation shows that
 $$\lim_{y \ra x}\frac{d_H(x, y)}{|x - y|_x} \geq 1.$$
{ Parts 1 and 2 of Lemma~\ref{l:1n} follow from Theorems~\ref{1tself} and \ref{1thyp}. Part 3 is a special case of part 2. Part 4 follows from adding up parts 1 to 3.}
\end{proof}
\begin{theorem}[Lov\'{a}sz, \cite{Lovasz}]\label{1thm:l}
Let $S_1$ and $S_2$ be measurable subsets of $K$. Let $\mu$ be the uniform probability measure on $K$. Then, \beqs \mu(K
\setminus (S_1 \cup S_2))  \geq \left(e^{d_H(S_1, S_2)} - 1\right)\mu(S_1)
\mu(S_2). \nonumber\eeqs
\end{theorem}
\begin{theorem}\label{1thmche}
If $F$ is a self-concordant barrier of $K$ with complexity parameter $\nu$, presented in the format of Section~\ref{ss:pres}, then
$$\C = \Omega\left(\frac{1}{\sqrt{n\nu}}\right).$$
\end{theorem}
\begin{proof}
Theorem~\ref{1thmche} follows from Theorem~\ref{1thm:l} and Lemma~\ref{l:1n}.
\end{proof}



%
%

\section{Analysis of the mixing time}\label{sec:analysis}
\subsection{Preliminaries}
We denote the conditional distribution of $x_{i+1}$ given $x_i = x$ by $P_x$.
 Lemma~\ref{1lnice}  is a statement about the concentration of derivatives of odd order  in high dimension. It will be used in the proof of Lemma ~\ref{1lem:gp}, which relates the Riemannian distance between two points $x$ and $y$ to the total variation distance between $P_x$ and $P_y$. Lemma~\ref{2lnice} states that if the unit Dikin ellipsoid around a point contains the unit ball, then the points at which a random line through $x$ chosen from the distribution induced by the uniform measure on the unit sphere intersects the boundary are, with high probability, at a distance $\Omega^*(\sqrt{n})$ from $x$. 
 \begin{lemma}[Concentration bound]\label{1lnice}
     Let $h$ be chosen uniformly at random from the unit sphere $S^n = \{v \,|\,\|v\| = 1\}$. Then, for any odd $k$, $$\p\left[D^kF(x)[h, \dots, h] > k\eps\sup_{\|v\|_x \leq 1} D^k F(x)[v, \dots, v]\right] \leq \exp\left(\frac{- n \eps^2}{2}\right).$$
    If $F$ is a self-concordant barrier, and $$\forall v,\, \|v\|^2 \geq D^2F(x)[v, v]$$   when $k = 3$, this simplifies to
    $$\p\left[D^3F(x)[h, h, h] > 3\eps\right] \leq \exp\left(\frac{- n \eps^2}{2}\right).$$
        \end{lemma}
\begin{proof}The ``Bernstein inequality" of Gromov (Section 8.5,  \cite{gro1})  which applies to multivariate polynomials restricted to $S^n$, states that for any polynomial $p$ on $S^n$ of degree $k$,
        $$\sup_{h \in S^n}\|\mathrm{grad}\, p(h)\| \leq k \sup_{h \in S^n} |p(h)|.$$
         For any fixed $x$, $D^kF(x)[h, \dots, h]$ is a polynomial in $h$ of degree $k$.
         Therefore $$\frac{D^kF(x)[h, \dots, h]}{k\sup_{\|v\|_x \leq 1} \big| D^k F(x)[v, \dots, v]\big|}$$ is 1-Lipschitz on $S^n$. If $k$ is odd, $D^kF(x)[h, \dots, h] = - D^kF(x)[-h, \dots, -h]$, and therefore its median with respect to the uniform measure $\sigma$ on the unit sphere is $0$. The first part of the lemma follows from the measure concentration properties of Lipschitz functions on the sphere   (page $44$ in \cite{Ball1});
         namely, if $f$ is an $1$-Lipschitz function on the unit sphere and $M$ is its median, then
         \beq\lab{eq:ball} \sigma\left(p > M +  \eps\right) \leq  e^{- \frac{n \eps^2}{2}}.\eeq When $F$ is a self-concordant barrier, the second statement follows because
    $$\sup_{\|v\| \leq 1} D^3 F(x)[v, v, v] \leq \sup_{\|v\|_x \leq 1} D^3 F(x)[v, v, v] \leq 1.$$
        \end{proof}

\begin{lemma}\lab{2lnice}
Let $P$ be a polytope and $x$ a point in it. Let the Dikin ellipsoid at $x$ with respect to the logarithmic barrier at $x$ contain the unit ball. Let $v$ be chosen uniformly at random from the unit ball centered at $x$ and $\ell$ be the line through $x$ and $v$, and $p$ and $q$ be the two points of intersection of $\ell$ with the boundary $\partial P$. Then, for any constant $\aaa \geq 0$, \beqs \p\left[\min(\|p\|, \|q\|) \leq  \sqrt{\frac{n}{2(\aaa +  2\ln n)}}\right] \leq  2 e^{-\aaa}. \eeqs
\end{lemma}
\begin{proof}
Without loss of generality, we may assume $x$ to be the origin. The unit ball is contained in the Dikin ellipsoid  and so $P$ can be expressed as $\bigcap_{i=1}^m \{a_i^T x \leq 1\}$, where \beq\lab{eq:simp}I \succeq \sum_{i=1}^m a_i a_i^T.\eeq Examining the trace and the norm on both sides of (\ref{eq:simp}), we obtain $$\forall i, \|a_i\| \leq 1$$ and $$\sum_{i=1}^m \|a_i\|^2 \leq n.$$
We note that
\beqs \min(\|p\|, \|q\|) = \left(\min_i |a_i^T x|\right)^{-1}. \eeqs
Thus, it is sufficient to show that
\beqs \p\left[\exists_i (a_i^T v)^2 \geq \frac{2(\aaa +  2\ln n)}{ n }\right] \leq 2 e^{-\aaa},\eeqs which we proceed to do.
Let $S$ be the subset of $[m] := \{1, \dots, m\},$ consisting of those $i$ for which $\|a_i\| \geq \frac{1}{\sqrt{n}}$. Clearly, if for some $i$,
$(a_i^T v)^2 \geq \frac{2(\aaa +  2\ln n)}{ n }$, then $i \in S$.
By (\ref{eq:simp}), $|S| \leq n^2$.
Thus, by the union bound,
{ \beqs \p\left[\exists_{i\in S} (a_i^T v)^2 \geq \frac{2(\aaa +  2\ln n)}{ n }\right] & \leq &  n^2 \sup_{i \in S} \p\left[(a_i^T v)^2 \geq \frac{2(\aaa +  2\ln n)}{ n }\right].\eeqs}
We note that, by (\ref{eq:ball}), for any vector $w$ with norm less or equal to $1$,
\beqs \p\left[a_i^T w \geq \sqrt{\frac{2(\aaa +  2\ln n)}{ n }}\right] \leq  \frac{e^{-\aaa}}{n^2},\eeqs
and so
\beqs \p\left[\exists_i (a_i^T v)^2 \geq \frac{2(\aaa +  2\ln n)}{ n }\right] \leq 2 e ^{-\aaa}. \eeqs
\end{proof}

In order to obtain mixing time bounds, we will first prove in Lemma~\ref{1lem:gp} that if two points $x$ and $y$ are nearby in that $d(x, y) \leq O(\frac{1}{\sqrt{n}})$, then the total variation distance between the corresponding distributions $P_x$ and $P_y$ is $< 1 - \Omega(1)$.

 For $x \neq y$, \beqs 1 - d_{TV}(P_x, P_y) =  \E_z\left[ \min\left(1, \frac{G_y(z)}{G_x(z)}, \frac{G_z(x)}{G_x(z)}, \frac{G_z(y)}{G_x(z)}\right)\right],\eeqs
    where the expectation is taken over a random point $z$ from the density $G_x$ and $$\min\left(1, \frac{G_y(z)}{G_x(z)}, \frac{G_z(x)}{G_x(z)}, \frac{G_z(y)}{G_x(z)}\right)$$ is defined to be $0$ if $z \not\in K$.

We will use the following fact (see Section $2.2$,  \cite{Nem_slides}) with $D^k F$ in the place of $M$.
\begin{fact}\label{1f1}
Let $M[h_1, \dots, h_k]$ be a symmetric $k$-linear form on $\RR^n$. Then,
$$M[h_1, \dots, h_k]  \leq  \|h_1\| \|h_2\| \dots \|h_k\| \sup_{\|v\| \leq 1} M[v, \dots, v].$$
\end{fact}

\begin{fact}\lab{1f2}
Let the eigenvalues  of the  covariance matrix of  an $n-$dimensional Gaussian $g$ be bounded above by $\lambda$. Let $<\cdot, \cdot>$ be an inner product and $v \in \RR^n$
Then, $\E[<v, g>^2] \leq \lambda <v, v>$.
\end{fact}

\subsection{Relating the Markov Chain to the manifold}

In this section, we find an upper bound on the total variation distance between the distributions of one-step transition probabilities corresponding at two points at a Riemannian distance of $O(1/\sqrt{n})$. We also describe two results used in the proof of Theorem~\ref{1tmain1}.
We will frequently make statements of the form
$$\p[g(x) > O(f(x))] \leq c.$$
By this we mean, there exists a universal constant $C$ such that
$$\p[g(x) > C f(x)] \leq c.$$

Finally, we will frequently make use of the facts from (Theorem 2.1.1, \cite{NNbook}) stated  below that  Dikin ellipsoids vary smoothly, and that they are contained in the convex set.
\begin{itemize}
\item Given any self-concordant barrier $F$, for any $y$ such that $$D^2F(x)[x - y, x - y] = r^2 < 1,$$
 for any vector $h \in \RR^n$,
 \beq \lab{eq:dikinfund}\,\,\,\,(1 - r)^2 D^2F(x)[h, h] \leq D^2F(y)[h, h] \leq \frac{1}{(1 - r)^2}D^2F(x)[h, h]. \eeq
 \item  The Dikin ellipsoid centered at $x$, having
radius $1$, is contained in $K$.
\end{itemize}
For two probability distributions $P_x$ and $P_y$, let $d_{TV}(P_x, P_y)$ represent the total variation distance between them.
Without loss of generality, let $x$ be the origin $0$ (which is achievable by translation), and for any $v$, let $D^2F(0)[v, v] = \|v\|^2$ (which is achievable by an affine transformation of $K$).

\begin{lemma}[Relating $d$ to Markov Chain]
\label{1lem:gp}
If $x, y \in K $ and $d(x, y) \leq
\frac{ 1}{C \sqrt{n}}$, then $d_{TV}(P_x, P_y) = 1 -  \Omega(1)$.
\end{lemma}

\begin{proof}
Without loss of generality, we may assume that ${F_h}$, ${F_\ell}$ and ${F_s}$ are strictly convex. In case any one is not, we can add the strictly convex logarithmic barrier of a sufficiently large cube, thereby making an arbitrarily small change to its second, third and if it is not ${F_s}$, fourth order derivatives uniformly over $K$.
Due to affine invariance, without loss of generality, let $<u, v>_x := <u, v>$, the usual dot product. As defined in Section~\ref{s:walk}, for any $z \in K$, $$V(z) = \frac{1}{2}\det D^2 F.$$
 By Lemma~\ref{1lNT}, it suffices to prove that there is an absolute constant $C$  such that if $x, y \in K $ and $\|x -y\|_x \leq
\frac{ 1}{C \sqrt{n}}$, then $d_{TV}(P_x, P_y) = 1 -  \Omega(1)$.
Without loss of generality, we assume $x$ is the origin and we drop this subscript at times to simplify notation.
\beqs 1 - d_{TV}(P_x, P_y) = \E_z\left[ \min\left(1, \frac{G_y(z)}{G_x(z)}, \frac{G_z(x)}{G_x(z)}, \frac{G_z(y)}{G_x(z)}\right)\right],\eeqs
    where the expectation is taken over a random point $z$ having density $G_x$. Thus, it suffices to prove the existence of some absolute constant $c$ such that
    $$\p\left[\min\left(\frac{G_y(z)}{G_x(z)}, \frac{G_z(x)}{G_x(z)}, \frac{G_z(y)}{G_x(z)}\right) > c\right] = \Omega(1).$$
This translates to
\begin{eqnarray*}\lab{longeqn}\p\left[\max\left( n\|y - z\|_y^2 - r^2V(y), n\|z\|_z^2 - r^2V(z),  n\|z - y\|_z^2 - r^2V(z)\right) -  n\|z\|^2 <  O(r^2)\right]\\ = \Omega(1).\end{eqnarray*}

Now we use the following lemmas whose proofs are in the Appendix.
\begin{lemma} \lab{lem:6mar28}$- V(y) < O(1)$ \end{lemma}
\begin{lemma} \lab{lem:7mar28}\beq \label{1eq11111} \p\left[ - V(z) < O(1) \right] & > &  \frac{9}{10}.\eeq
\end{lemma}

\begin{lemma}\lab{prop:mar21-14}
\beqs\p\left[\max\left(\|y - z\|_y^2,  \|z \|_z^2 ,  \|z - y\|_z^2 \right) - \|z\|^2 < O(\frac{1}{n}) \right] > \frac{199}{1000}.\eeqs
\end{lemma}
Substituting the statements of the last three lemmas into the preceding expression of probability completes the proof.
\end{proof}
The following two results are needed in the proof of Theorem~\ref{1tmain1}.
Lemma~\ref{1lcond} is proved in the Appendix { and the proof is based on that of a theorem in \cite{Lovasz}.}
\begin{lemma}[Bound on Conductance]\label{1lcond}
Let $\mu$ be the uniform distribution on $K$.
The conductance $$\Phi := \inf\limits_{\mu(S_1) \leq \frac{1}{2}}\frac{\int_{S_1} P_x(K\setminus S_1) d\mu(x)}{\mu( S_1)}$$ of the Markov Chain in Dikin Walk Algorithm  is $\Omega(\C)$.
\end{lemma}

\begin{theorem}[Lov\'{a}sz-Simonovits \cite{LS1}]\label{1thm:L1}
Let $\mu_0$ be the initial distribution for a lazy reversible
ergodic Markov Chain whose conductance is $\Phi$ and stationary
measure is $\mu$, and $\mu_k$ be the distribution of the $k^{th}$
step. Let $M:=\sup_S \frac{\mu_0(S)}{\mu(S)}$ where the supremum is
over all measurable subsets $S$ of $K$. For every bounded $f$, let $\|f\|_{2, \mu}$ denote
$\sqrt{\int_K f(x)^2 d\mu(x)}$. For any fixed $f$, let $Ef$ be the
map that takes $x$ to $\int_K f(y) dP_x(y)$. Then, \ben \item for all $S$,
$$|\mu_k(S) - \mu(S)| \leq \sqrt{M}\left(1 -
\frac{\Phi^2}{2}\right)^k.$$ \item If $\int_K
f(x) d\mu(x) = 0$,
$$\|E^k f\|_{2, \mu} \leq \left(1 - \frac{\Phi^2}{2}\right)^k \|f\|_{2, \mu}.$$\een
\end{theorem}

\subsection{Proof of Theorem~\ref{1tmain1}}\lab{ssec:1tmain1}
\begin{proof}[Proof of Theorem~\ref{1tmain1}]
We first give a summary of the proof.
By Theorem~\ref{1thmche},  $\beta_{fat}$ is bounded below by $\Omega\left(\frac{1}{\sqrt{n\nu}}\right)$. Lemma~\ref{1lcond} states that the conductance $\Phi$ of the Markov Chain is bounded below by $\Omega\left(\beta_{fat}\right)$. We  substitute this in Theorem~\ref{1thm:L1} to recover the fact that the mixing time from a warm start is $O\left(n\nu\ln \frac{1}{\eps}\right)$.

To recover the fact that the mixing time from an ``$s-$central" point is $$O\left(n\nu\left(n \ln (\nu s) + \ln \frac{1}{\eps}\right)\right),$$ We first get a bound of $\exp(n \ln (\nu s))$ on the $\LL_2$ norm of the starting density, which we then use in Theorem~\ref{1thm:L1}. This concludes our summary -- we now proceed with the proof.


Let $e$ be the first time the Markov Chain escapes $x_0 = 0$. Thus $e$ is an integer valued random variable defined by the event that $y_e$ is the first point in $y_1, \dots$ that is not equal to $y_0$. Let the density of $y_e$ be $\rho_e$.  We know that $\forall t$, $\p[e \geq t+1 |e \geq t] \leq 1 - \Omega(1)$,  therefore
\beqs \sup_x \frac{\rho_e(x)}{G_0( x)} & = & O(1).\eeqs
{ Let $\mu(S)$ denote the measure assigned by the uniform probability measure on $K$ on a measurable set $S$.
Therefore \beqs \int_{K} \left(\vol(K) \rho_e(x)\right)^2 d\mu(x) & = &  \exp\left(O(n \ln (s \nu))\right). \eeqs
To see the last step, consider a linear transformation that maps the Dikin ellipsoid at the origin to the unit ball. Then the convex set is contained inside a ball of radius $O(s\nu)$, making $\left(\vol (K)\right)$ bounded above by $\left(\frac{O(s^2\nu^2)}{n}\right)^{n/2}$  because the volume of a  ball in $n-$dimensions of radius $R$ is $\left(\frac{O(R^2)}{n}\right)^{n/2}$.  Also, $\rho_e$ is bounded above in sup-norm by
$\exp((n/2)\ln(O(n)))$, because $G_0(x)$ is so bounded.  Therefore $$\int_K \rho_e(x)^2 dx < \sup_K \rho_e(x) \int_K \rho_e(x) dx < (O(n))^{n/2}.$$   This concludes our explanation.}

For every bounded $f$, let $\|f\|_{2, \mu}$ denote
{ $\sqrt{\int_K (\vol(K) f(x))^2 d\mu(x)}$.}
Let $f_e(x):= \rho_e(x) - (\vol(K))^{-1}.$
We then see from the above computation that {\beq \|f_e\|_{2, \mu} \leq \|\rho_e(x)\|_{2, \mu} + \|\left(\vol (K)\right)^{-1}\one_K\|_{2, \mu} < \exp\left(O(n \ln ( s \nu))\right). \label{eq:L2}\eeq}

 Theorem~\ref{1thm:L1} states that $$\forall S, |\mu_k(S) - \mu(S)| \leq \sqrt{M}\left(1 -
\frac{\Phi^2}{2}\right)^k.$$ Therefore, if $$k > \frac{\ln \sqrt{\frac{M}{\eps^2}}}{- \ln \left(1 - \frac{\Phi^2}{2}\right)} = O\left(\ln \left(\frac{1}{\eps}\right) n \nu\right),$$ then,
$$\forall S, |\mu_k(S) - \mu(S)| \leq \eps.$$
This completes the proof of mixing from a warm start { \ie when  $M= O(1)$.}

To recover the proof of mixing from a $s-$central point, we see by Theorem~\ref{1thm:L1}, Lemma~\ref{1lcond} and (\ref{eq:L2}) that
\begin{eqnarray} \|E^k f_e\|_{2, \mu} & \leq &  \left(1 - \frac{\Phi^2}{2}\right)^k \|f_e\|_{2, \mu}\\
                                      & \leq & \left(1 - \Omega(\frac{1}{n\nu})\right)^k \exp\left(O(n \ln (s  \nu))\right). \end{eqnarray}
                                      Therefore
                                      for $k = \Omega(n\nu(\ln(1/\eps) + n \ln (s  \nu)))$, we have  $\|E^k f_e\|_{2, \mu} \leq \eps$.
This completes the proof of Theorem~\ref{1tmain1}.
\end{proof}

\section{Mixing in a direct product of convex sets}\label{sec:direct}
\subsection{Preliminaries}

Our analysis of mixing in a direct product of convex sets in Theorem~\ref{thm:bob} hinges upon a lower bound on the Cheeger constant $\C$, which is obtained by comparing the isoperimetry of the weighted manifold $\MM$ obtained by equipping $K$ with the metric from the Hessian of $F$, with the isoperimetry of the Dikin metric on the $h-$dimensional cube $[-\frac{\pi}{2}, \frac{\pi}{2}]^h$, with respect to the  barrier $F_\square(x_1, \dots, x_h) := - \sum_i \ln \cos(x_i)$ (see Section 6.2, \cite{NT}).

For a manifold $\NN$ equipped with a measure $\mu$ and metric $d$, let the Minkowski outer measure of a (measurable) set $A$ be defined as
 \beqs \mu^+( A) := \liminf_{\eps \ra 0^+} \frac{\mu(A_\eps) - \mu(A)}{\eps}, \eeqs
 where $A_\eps := \{x|\dm(x, A) < \eps\}.$
\begin{definition}\lab{def:cheeger}
The (infinitesimal)  Cheeger constant of the weighted manifold $(\NN, \mu)$ is
\beqs \beta_\NN = \inf_{A \subset \MM, \mu(A) \leq \frac{1}{2}} \frac{\mu^+(A)}{\mu(A)}, \eeqs where the infimum is taken over measurable subsets. \\
The isoperimetric function of $\mu$ is the largest function $I_\mu$ such that  $\mu^+(A) \geq I_\mu(\mu(A))$ holds for all Borel sets.
\end{definition}
Let the $h-$fold product space $\NN \times \dots \times \NN$ be denoted $\NN^h$, where the distance between points $(x_1, \dots, x_h)$ and $(y_1, \dots, y_h)$ is $\sqrt{\sum_i d(x_i, y_i)^2}$.

We will need the following theorem of Bobkov and Houdr\'{e} (Theorem 1.1 \cite{bobkov}).
\begin{theorem}\lab{thm:bob2}
For any triple $(\NN, d, \mu)$ as above,
\beq \lab{eq:bob1} I_{\mu^{\otimes h}} \geq \frac{1}{2\sqrt{6}} I_\mu. \eeq
\end{theorem}

We will also need the following theorem, which is a  modification of Barthe (Theorem 10, \cite{barthe}), obtained by scaling the metric on $\R$ by $\frac{1}{\sqrt{\nu}}$.

\begin{theorem}\lab{thm:bart}
Let $k \geq 2$ be an integer. For $i = 1, \dots, k,$ let $(X_i, d_i, \mu_i)$ be a Riemannian manifold, with its geodesic distance and an absolutely continuous Borel measure of probability and let $\la_i$ be a probability measure on $\R$ with even log-concave density. If $I_{\mu_i} \geq \frac{ I_{\la_i}}{\eta}$ for $i = 1, \dots, k,$ then
$$I_{\mu_1 \otimes \dots \otimes \mu_k} \geq \frac{I_{\la_1 \otimes \dots \otimes \la_k}}{\eta}.$$
\end{theorem}



The following lemma allows us to relate the ``fat" Cheeger constant $\C$ with the infinitesimal version $\beta_\M$.
\begin{lemma}\lab{l:man1}
Let $\AAA, \BBB \subseteq \MM$ and $d_\M(\AAA, \BBB) \geq \de_\M$. Then,
\beqs \mu(\MM \setminus \{\AAA \cup \BBB\})  \geq  2\min(\mu(\AAA), \mu(\BBB))(e^{\frac{\beta_\M \de_\M}{2}}  - 1).\eeqs
\end{lemma}
\begin{proof}
We will consider two cases.

First, suppose that $\max(\mu(\AAA), \mu(\BBB)) < \frac{1}{2}$.
Without loss of generality, we assume that $\mu(\AAA) \leq \mu(B).$
Then, let $$\delta_1 := \sup\limits_{\mu(\AAA_\de) < \frac{1}{2}} \de.$$
We proceed by contradiction to show that $$ \mu(\MM\setminus\{\AAA \cup \BBB\})  \geq  2\mu(\AAA)\left(e^{\frac{\beta_\M\de_\M}{2}} - 1 \right).$$
 Suppose for some $\beta < \beta_\M$,
\beq\exists \de \in [0, \de_1), \mu(\AAA_\de) < e^{\beta\de} \mu(\AAA).\lab{44}\eeq
Let $\de'$ be the infimum of such $\de.$ Note that 
since $\mu(\AAA_\de)$ is a monotonically increasing  function of $\de$,
$$\mu(\AAA_{\de'}) \geq e^{\beta \de'} \mu(\AAA).$$
However, we know that
\beqs \mu^+(\partial \AAA_{\de'}) := \liminf_{\eps \ra 0^+} \frac{\mu(\AAA_\eps) - \mu(\AAA)}{\eps} \geq \beta_\M,\eeqs
which contradicts the fact that in any right neighborhood of $\de'$, there is a $\de$ for which (\ref{44}) holds.
This proves that for all $\de \in [0, \de_1)$,\, $\mu(\AAA_\de) \geq e^{\de\beta_\M} \mu(\AAA).$
We note that $\AAA_{\de_1} \cap \BBB_{\de_\M - \de_1} = \emptyset$,  therefore $\mu(\BBB_{\de_\M - \de_1}) \leq \frac{1}{2}$. So the same argument tells us that
\beq \mu(\BBB_{\de_\M - \de_1}\setminus \BBB)\geq (e^{\beta_\M(\de_\M - \de_1)}-1)\mu(\BBB).\lab{14}\eeq
Thus,
$ \mu(\MM\setminus\{\AAA \cup \BBB\})  \geq  \mu(\AAA)(e^{\beta_\M(\de_\M - \de_1)} + e^{\beta_\M\de_1} - 2). $
This implies that
$$ \mu(\MM\setminus\{\AAA \cup \BBB\})  \geq  2\mu(\AAA)\left(e^{\frac{\beta_\M\de_\M}{2}} - 1 \right).$$

Next, consider the second case; suppose $\mu(\BBB) \geq \frac{1}{2}.$ We then set $\de_1 := \de_\M$, and see that the arguments from (\ref{44})  to (\ref{14})
carry through verbatim. Thus, in this case, $ \mu(\MM\setminus\{\AAA \cup \BBB\})  \geq  \mu(\AAA)(e^{\beta_\M\de_\M}  - 1)  \geq 2\mu(\AAA)\left(e^{\frac{\beta_\M\de_\M}{2}} - 1 \right).$

\end{proof}

This immediately leads to the following corollary:
\begin{corollary}\lab{cor:imp1}
$$\C = \Omega\left(\frac{\beta_\M}{\sqrt{n}}\right).$$
\end{corollary}
Nesterov and Todd show in (Lemma 4.1, \cite{NT}) that the Riemannian metric $\MM$ on the direct product of convex sets induced by $\sum_{i = 1}^h F_i$ is the same as the direct product of the Riemannian metrics $\MM_i$ induced by individual $F_i$ on the respective convex sets $K_i$.

Theorem~\ref{thm:bob} illustrates the utility of using the Riemannian metric in proving isoperimetric bounds in one specific case, namely when the convex set of interest is a product of convex sets having smaller dimension. A large number of combinatorial optimization questions can be viewed as optimizing a non-convex quadratic function over a unit cube. This is one scenario where we hope the improved bounds on the mixing time may lead to interesting results in the future.

\subsection{Proof of Theorem~\ref{thm:bob}}\begin{proof}In order to show that the mixing time is $O(n\nu)$, by Theorem~\ref{1thm:L1}, it suffices to show that the conductance $\Phi$ is bounded below by
$\Omega(\frac{1}{\sqrt{n\nu}}).$ By Lemma~\ref{1lcond}, it is in turn enough to bound $\beta_{fat}$ from below by $\frac{1}{\sqrt{n\nu}}$. By Corollary~\ref{cor:imp1}, it suffices to show that $\beta_\M = \Omega(\frac{1}{\sqrt{\nu}}).$
We will show this using Theorem~\ref{thm:bart} and Theorem~\ref{thm:bob2}.
Consider the $h-$dimensional cube $[-\frac{\pi}{2}, \frac{\pi}{2}]^h$, and the metric from the Hessian of the  barrier $F_\square(x_1, \dots, x_h) := - \sum_i \ln \cos(x_i)$ (see Section 6.2, \cite{NT}). The map
$$\psi:(x_1, \dots, x_h) \mapsto (\ln(\sec(x_1) + \tan(x_1)), \dots, \ln(\sec(x_h) + \tan(x_h))),$$ maps the cube with the Hessian metric isometrically onto Euclidean space, and the push-forward of the uniform density on the cube is a density  $\phi \otimes \dots \otimes \phi$ on $\R^h$, where
 $$\phi(x) = \frac{ \cos\left(2 \arctan\left(\frac{e^{x} - 1}{e^x + 1}\right)\right)}{\pi},$$ and $$\frac{d^2 \ln \phi}{dx^2} = -\frac{4 e^{2x}}{(1 + e^{2x})^2} < 0,$$ and the density is even (thus meeting the conditions of Theorem~\ref{thm:bart}). In the $1-$dimensional case, it is easy to check that the barrier is $1-$self-concordant (Section 6.2 \cite{NT}). Therefore, in the $1-$dimensional case, $I_\phi$ is bounded above and below by fixed constants. This, together with Theorem~\ref{1thmche} implies that for each $\MM_i$ and the uniform measure  $\mu_i$ (on $K_i$),  the isoperimetric profile $I_{\mu_i}$ of $(\MM_i, \mu_i)$
satisfies $I_{\mu_i} \geq \Omega(\frac{1}{\sqrt{\nu}}){I_\phi}$.
Now applying Theorem~\ref{thm:bart} and Theorem~\ref{thm:bob2} in succession,
we see that
$$I_{\mu_1 \otimes \dots \otimes \mu_h} \geq  \Omega\left(\frac{1}{\sqrt{\nu}}\right)I_{\phi \otimes \dots \otimes \phi} \geq  \Omega\left(\frac{1}{\sqrt{\nu}}\right)I_{\phi} = \Omega\left(\frac{1}{\sqrt{\nu}}\right){\mathbf{1}},$$ where ${\mathbf{1}}$ is the constant function taking the value $1$.
Therefore,
$$\beta_\M = \Omega(\frac{1}{\sqrt{\nu}}).$$
\end{proof}

\section{Analysis of Las Vegas Algorithm}\label{sec:convex}
\begin{proof}[Proof of Theorem~\ref{1tmain2}]
 Note that this Markov Chain is not ergodic and has no stationary probability distribution.  We will analyze its behavior up to time $t$ by relating this to the limiting behavior of  Dikin walks on a family of convex sets $\{\tk^j\}_{j \geq 1}$ each contained in the next, such that $ \tk^j = \tk \cap \{x|c^Tx \leq j\}$. Note that for any fixed $j$, our mixing results from Theorem~\ref{1tmain1} apply since $\tk^j$ is bounded. Let $\tf_j = \tf + \ln (j - c^T x)$. By known properties of barriers (\cite{NNbook}), the self-concordance parameter of $\tf_j$ is at most $1$ more than that of $\tf$. As $j$ tends to $\infty$,  $D^2 \tf_j$ converges uniformly to $D^2 \tf$ in the $\ell_2^n \ra \ell_2^n$ operator norm on Hessian matrices on any compact subset of $\tk$.
Therefore, for any $t$, the distribution of the $t$-tuple $(T(x_0), T(x_1), \dots, T(x_t))$ is the limit in total variation distance of  the distributions of $t$-tuples $(T(x^j_0), T(x^j_1), \dots, T(x^j_t))$, where $(x^j_0, \dots, x^j_t)$ is a random walk on $\tk^j$ starting at $0$.

We will now give an upper bound for $\p[c^T(x_i) \leq 1 - \eps]$.
%
%
Let $y_i := T(x_i)$.
Let $e$ be the first time the Markov Chain $y_0, y_1, \dots$ escapes $y_0$. Thus $e$ is an integer valued random variable defined by the event that $y_e$ is the first point in $y_1, \dots$ that is not equal to $y_0$. Let the density of $y_e$ be $\rho_e$. $\forall t$, we know from Lemma~\ref{1lem:gp} (applied when $y \ra x$ ) that  $\p[e \geq t+1 |e \geq t] \leq 1 - \Omega(1)$,  therefore
\beqs \sup_x \frac{\rho_e(x)}{\tg_0( x)} & = & O(1).\eeqs
Without loss of generality, in the rest of this proof, we assume that for all $v$, $D^2\tf(0)[v, v] = \|v\|^2.$

Therefore { \beq \int_{\tk} \rho_e(x)^2 dx & = & O(\int_{\RR^n} \tg_0(x)^2 dx)\\
& = & {\exp\left(O(n \ln (n ))\right)}. \lab{c12apr4}\eeq}
Let $\rho_t(x)$ be the density of $x_t$. Then, by Theorem~\ref{1tmain1} and the fact that as far as total variation distance is concerned, a random walk on $\tk$ can be viewed as the limit of random walks on the $\tk^j$ as $j \ra \infty$,
{\beq \frac{\int_{\tk} \rho_t(x)^2 dx}{\int_{\tk} \rho_e(x)^2 dx} = O((1 - \frac{\Phi^2}{2})^t).\lab{c13apr4}\eeq}
{ By Lemma~\ref{1lcond}, this is less than
$\exp\left(- t \Omega(\C^2)\right)$, using the fact that $1 - x \leq e^{-x}$.

The following lemma is proved in the Appendix.}
\begin{lemma}\label{1lem:l2}
Let $\tk_\eps = T(K \cap \{c^T x \leq 1-\eps\})$. Let $\rho$ be a density supported on $\tk$ such that $$\int_{\tk_\eps} \rho(x) dx \geq \de.$$ Then,
\beqs \int_{\tk} \rho(x)^2 dx
& \geq &  \de^2 \exp\left(- O\left(n \ln \left(\frac{s \nu}{\eps}\right)\right)\right).\eeqs
\end{lemma}

{ Using (\ref{c12apr4}), (\ref{c13apr4}),  and Lemma~\ref{1lem:l2}, if $ \p\left[c^T x_t \leq 1 - \eps\right] = \delta$, we have that there exists an  absolute constant $C>0$ such that}
{ $$\frac{\de ^2 \exp\left(-C\left(n \ln \left(\frac{s\nu}{\eps}\right)\right)\right)}{\exp(C(n \ln n))} < e^{- t (\beta_{fat})^2/C}.$$}
Therefore, { $$\p\left[c^T x_t \leq 1 - \eps\right] \leq \exp\left(C\left(n \ln \left(\frac{n s \nu}{\eps}\right)\right) - (t (\C)^2)/C \right).$$} This gives the desired upper bound on $\p\left[c^T x_t \leq 1 - \eps\right]$. We next proceed to get an expression for $\tau(\eps, \de)$.
We have
 {\beqs\p\left[(\exists t \geq \tau) \, c^T x_\tau \leq 1 - \eps\right] & \leq & \sum_{t \geq \tau} \exp\left(C\left(n \ln \left(\frac{n s \nu}{\eps}\right)\right) - t( \C^2)/C \right)\\
& \leq & \frac{\exp\left(C\left(n \ln \left(\frac{n s \nu}{\eps}\right)\right) - \tau (\C^2)/C \right)}{1 - \exp\left(- \left(\C\right)^2/C\right)}.\eeqs}
Therefore, for any $\de$, $\p\left[(\exists t \geq \tau) \, c^T x_\tau \leq 1 - \eps\right] < \de$ for
 { $$\tau >  C'\left(\frac{1}{(\C)^2} \left( \ln \frac{1}{\de} + \left(n \ln \frac{n s \nu}{\eps}\right)\right)\right),$$ for a universal constant $C'$.  Together with the fact that $\beta_{fat} = \Omega\left(\frac{1}{\sqrt{n\nu}}\right)$ (from Theorem~\ref{1thmche}) this completes the proof.}
\end{proof}

\section{Implementing the barrier oracle in the linear and semidefinite cases}\lab{sec:lin_sem}

The most frequently encountered barrier functions are the logarithmic barrier for polytopes and the $\log\det$ barrier for convex sets defined by semidefinite constraints (See Section~\ref{secdefn}).
We discuss the implementation of the barrier oracle for the logarithmic barrier below, in the case where $x$ is in the set.
Let  $K_\ell$ be the set of points satisfying the system of inequalities
$Ax \leq \one$. Then, $H(x) = A^T D(x)^2 A$ where $D(x)$ is the diagonal matrix whose $i^{th}$ diagonal entry $d_{ii}(x) = \frac{1}{1 - a_i^Tx}$ .\\

By results of Baur and Strassen \cite{BS},
the complexity of solving linear equations and of computing the
determinant of an $n \times n$ matrix is
$O(n^\gamma)$ where $\gamma < 2.377$
is the exponent for matrix multiplication.  The computation of $A^T D(x)^2 A$ can be achieved using $mn^{\gamma-1}$ arithmetic operations, by partitioning
 a padded extension of $A^T D$ into $\leq \frac{m + n - 1}{n}$ square matrices. Thus, the complexity of the barrier oracle is $O(m n^{\gamma-1})$
arithmetic operations.

In case the convex set is defined by a semidefinite constraint of rank $\nu$, the $\log\det$ barrier is a hyperbolic barrier and has a self-concordance parameter of $\nu$ and a {\it complexity parameter} (defined in \ref{eq:compl}) of $n\nu$.   the number of arithmetic steps needed for computing the Hessian of the $\log \det$ barrier is $O(n^2 \nu^2 + n \nu^\gamma)$, (see Section 11.3, \cite{Nem_slides}. We have replaced an exponent $3$ in \cite{Nem_slides} with $\gamma$).
Given the Hessian, it can be inverted in $(n^{\gamma})$ arithmetic steps. This is needed to implement one step of the Dikin walk.
Suppose $K$, as above, is a  convex set 
that is defined by $m$ linear constraints and additionally, semidefinite constraints of total rank $\nu_h$ (which can be as low as $O(1)$, e.\,g.\, for the intersection of a constant number of ellipsoids).
Then, the number of arithmetic steps for implementing one Dikin step is $$O\left(m  n^{\gamma - 1} + n^2 \nu_h^2 + n \nu_h^{\gamma}\right).$$
For Hit-and-Run, the number of arithmetic steps needed to make one move in   a naive implementation is $$O^*\left(\log(R/r)(mn + n\nu_h^2 + \nu_h^{\gamma})\right),$$ (since the natural way of certifying positive semidefiniteness is to take a Cholesky factorization, which has a complexity $O(\nu_h^{\gamma})$, computing the new semidefinite matrix after one step has a complexity $n(\nu_h^2)$ (Section 11.3, \cite{Nem_slides}) and testing containment in the region defined by linear constraints takes $O(nm)$ operations).  We see that
\ben
\item If $m < n \nu_h^2 + \nu_h^\gamma$, then the ratio between the number of arithmetic steps  for one move of Dikin walk and one move of Hit-and-Run is not more than $O^*(n)$. 
\item If $m \geq  n \nu_h^2 + \nu_h^\gamma$, then the ratio between the number of arithmetic steps  for one move of Dikin walk and one move of Hit-and-Run is not more than $O^*(n^{\gamma - 2}) < O^*(n^{0.38})$.
    \end{enumerate}
Combining the  arithmetic complexity of implementing one step of Hit-and-Run with the mixing time, the ratio between the number of arithmetic steps needed to produce one random point using Dikin walk to the number of arithmetic steps needed for producing one random point using Hit-and-Run is
$O^*\left(\frac{(m + \nu_h n) r^2}{R^2}\right)$ if $m < n \nu_h^2 + \nu_h^\gamma$ and
$O^*\left(\frac{(m + \nu_h n) r^2}{R^2 n^{0.62}}\right)$
if $m \geq n \nu_h^2 + \nu_h^\gamma$.

\subsection{Implementing one Dikin step}
 If $K$ is defined by semidefinite constraints, from a point $x \in K$, one step  for Hit-and-Run requires $\Omega(\log(R/d))$ membership operations, each of which requires testing the semidefiniteness of a $\nu \times \nu$ matrix (which takes $O(\nu^\gamma)$ arithmetic steps), where $R$ is the radius of a circumscribing ball, and $d$ is the distance of $x$ to the boundary of $K$.  Convex sets defined by semidefinite programs can be very ill-conditioned, and  the best possible a priori upper bound on $\log \frac{R}{d}$ is not less than $e^L$ where $L$ is the total bit-length of rational data defining $K$ and the point \cite{Ramana}.
In the general setting, the number of arithmetic operations needed for implementing a Dikin step would be independent of $R/r$, but would depend on two affine-invariant quantities - the parameter associated with the barrier and  $\log s$, where the starting point is $s-$central. In  ill-conditioned semidefinite programs, $\log s$ can be exponential in the bitlength, but for special points it can be much smaller; for example, for the center of mass and or the analytic center, it is $O(\log n)$ and $O(\log \nu)$ respectively.

\section{Concluding Remarks}

We  developed randomized analogues of barrier-based interior point methods, and  demonstrated their use in sampling convex sets and optimizing a linear function over a convex set. One potential application of these methods is to integer programming as shown by Huang and Mehrotra \cite{HuangMehrotra}.
It remains to be seen whether a more efficient algorithm for computing the volume of a polytope can be constructed using Dikin walk.
{
\section{Acknowledgements}
I thank Robert Freund and Ravi Kannan for stimulating conversations and Partha Niyogi for insights that motivated the view of Dikin walk as a random walk on a manifold. I thank the anonymous referee for a careful reading and many critical and insightful comments, which { I hope} have led to an improvement in the readability of this paper.}

\appendix
\section{Proofs of Lemmas}

\begin{proof}[Proof of Lemma~\ref{lem:6mar28}]
Let $\F := F - {F_\ell}$. For $w \in K$, let $\X(w) := D^2 {F_\ell}(w) - D^2 {F_\ell}(0)$.
By Lemma 12 in \cite{KNsample}, for any point $w \in K$,
{ the gradient of $\Tr \X$ at $w$ measured using $\|_\ell \cdot \|_w$ is $\leq 2 \sqrt{n}$. Therefore, the gradient of $\Tr \X$ at $0$ measured using $\|_\ell \cdot \|$ is $\leq O( \sqrt{n})$. $\X(0) = 0$, therefore,
$|\Tr \X(y)| \leq O(\|y\| \sqrt{n}) = O(1),$
since 
$\|y\| = O(1/\sqrt{n}).$ For the same reason,
 $\|X(y)\| = O( \sup_{\|w\| \leq \|y\|} \|D^3 {F_\ell}(w)[y, \cdot, \cdot]\|) = O(1/\sqrt{n})$.}

 For $w \in K$, let $\Y(w) := D^2 \F(w) - D^2 \F(0)$.
Then
{ \beqs\lab{eqmar28.1}\|\Y(y)\| = D^2\F(y) - D^2\F(0) = O( \sup_{\|w\| \leq \|y\|} \|D^3 \F(w)[y, \cdot, \cdot]\|) =O(1/n).\eeqs} Therefore, $|\Tr \Y(y)| = O(1).$
 \beqs - V(y) & = & - \ln  \det  (I + \X + \Y)\\
& = & - \Tr (\X + \Y  + R), \eeqs
where $R$ is a matrix whose $\|\cdot\| \ra \|\cdot\|$ norm $\|R\|$ is bounded above by $O(\max(\|X\|^2, \|Y\|^2)) = O(1)$.
  Thus, $|V(y)| = O(1),$ and Lemma~\ref{lem:6mar28} is proved.\end{proof}
\begin{proof}[Proof of Lemma~\ref{lem:7mar28}] { We need to show that \beqs \p\left[ - V(z) < O(1) \right] & > &  \frac{9}{10}.\eeqs}
 Let $\F := F - {F_\ell}$. For $w \in K$, let $\xx(w) := D^2 {F_\ell}(w)$, and let $\yy(w) := D^2 \F(w)$.
{ \beqs - V(z) & = &  \frac{-1}{2}\ln \det (\xx(z) + \yy(z))\nonumber\\
& = &  \frac{-1}{2}\left(\Tr \ln (\xx(z) + \yy(z))\right) \,\,\text{\wpot (\ie if $\|\xx(z) + \yy(z) - I\|<1$) }\nonumber\\
& = &  \frac{ -1}{2} \left(\Tr(\xx(z) - \xx(0)) + \Tr(\yy(z) - \yy(0))\right)\lab{eq:29-15}\\ & +&  \frac{1}{4}\left(\Tr\left(\xx(z) + \yy(z) - I\right)^2  \right)\lab{eq:29-16}\\ & - & O\left( \big|\Tr\left(I - (\xx(z) + \yy(z))\right)^3\big|\right) \,\,\,\,\,\text{ expanding by power series}\lab{eq:29-17}.
\eeqs}
{ This holds \wpot.}
The lemma is a consequence of the following lemmas:
\begin{lemma}\lab{cl:1}
{ $$\p\left[-(\Tr(\yy(z) - \yy(0))) \leq O(1)\right] \geq \frac{99}{100}.$$}
\end{lemma}
\begin{lemma}\lab{cl:2}
{ $$\p\left[\Tr(-(\xx(z) - \xx(0))) \leq O(1)\right] \geq \frac{99}{100}.$$}
\end{lemma}
\begin{lemma}\lab{cl:2.5}
{ The following two probabilistic inequalities hold.
\beqs \p\left[\big|\Tr\left(I - (\xx(z) + \yy(z))\right)^3\big| \leq O(1)\right] \geq \frac{99}{100}. \eeqs
\beqs \p\left[\big|\Tr\left(I - (\xx(z) + \yy(z))\right)^2\big| \leq O(1)\right] \geq \frac{99}{100}. \eeqs}
\end{lemma}

\end{proof}

\begin{proof}[Proof of Lemma~\ref{prop:mar21-14}]

Since $\|y\| < O(\frac{1}{\sqrt{n}})$ and $\|z\| < \frac{1}{3}$ with probability  greater than $1 - 10^{-3}$, $\|y\|_y$ and $\|y\|_z$ are $O(\frac{1}{\sqrt{n}})$.
So it suffices to show that
\beqs\p\left[\max\left(\|z\|_y^2 - \|z\|^2, <y, z>_y,  \|z \|_z^2 - \|z\|^2,   <y, z>_z \right)  < O(\frac{1}{n}) \right] & > & \frac{2}{10}.\eeqs
This fact follows from the following three lemmas and the union bound.
The proof of Lemma~\ref{l:ell} would go through if $\frac{4}{10}$ were replaced by $\frac{1}{2} - \Omega (1)$.
\begin{lemma} \lab{l:ell}\beqs\p\left[\max\left(\|_\ell z\|_y^2 - \|_\ell z\|^2, <_\ell  y, z>_y,  \|_\ell z \|_z^2 - \|_\ell z\|^2,   <_\ell  y, z>_z \right)  < O(\frac{1}{n}) \right] & > & \frac{4}{10}.\eeqs
\end{lemma}
\begin{lemma} \lab{l:h}\beqs\p\left[\max\left(\|_h z\|_y^2 - \|_h z\|^2, <_h y, z>_y,  \|_h z \|_z^2 - \|_h z\|^2,   <_h y, z>_z \right)  < O(\frac{1}{n^2}) \right] & > & \frac{9}{10}.\eeqs
\end{lemma}
\begin{lemma} \lab{l:s}\beqs\p\left[\max\left(\|_s z\|_y^2 - \|_s z\|^2, <_s y, z>_y,  \|_s z \|_z^2 - \|_s z\|^2,   <_s y, z>_z \right)  < O(\frac{1}{n^3}) \right] & > & \frac{9}{10}.\eeqs
\end{lemma}
\end{proof}

\begin{proof}[Proof of Lemma~\ref{1lcond}]
Let $S_1$ be a measurable subset of $K$ such that $\mu(S_1) \leq \frac{1}{2}$  and $S_2 := K \setminus S_1$ be its complement.
For any $x \neq y \in K$, \beqs
\frac{dP_y}{d\mu}(x) & = &
\frac{dP_x}{d\mu}(y).\nonumber\eeqs Let $S_1' = S_1 \cap \{x \big|
P_x(S_2) = o(1)\}$ and $S_2' = S_2 \cap
{ \{y \big|P_y(S_1) = o(1)\}}$. By the
reversibility of the chain, which is easily checked,
 $$\int_{S_1}  P_x(S_2) d\mu(x)  =  \int_{S_2}  P_y(S_1) d\mu(y).$$
If $x \in S_1'$ and $y \in S_2'$ then $$d_{TV}(P_x, P_y) := 1 - \int_K \min\left(
\frac{dP_x}{d \mu}(w),
\frac{dP_y}{d\mu}(w)\right)d\mu(w) = 1 - o(1).$$
 Lemma~\ref{1lem:gp} { states} that for an absolute constant $C$, if
$d(x, y) \leq
{ \frac{ 1}{C \sqrt{n}}}$, then $d_{TV}(P_x, P_y) = 1 -  \Omega(1)$. Therefore Theorem~\ref{1thmche}
implies that \beqs \mu((K \setminus S_1') \setminus S_2')  \geq \Omega(\C)
\min(\mu(S_1'), \mu(S_2')). \nonumber \eeqs First suppose $ \mu(S_1') \geq (1 - \Omega(1))
\mu(S_1)$ and $ \mu(S_2') \geq (1 - \Omega(1)) \mu(S_2)$. Then, \beqs
\int_{S_1} P_x(S_2) d\mu(x) & \geq & \mu((K \setminus
S_1') \setminus
S_2')\nonumber\\
& \geq & \Omega(\C)\mu(S_1') \nonumber\\
& \geq & \Omega(\C)\min(\mu(S_1'), \mu(S_2'))\qquad\nonumber\eeqs and we are done. Otherwise, without loss of
generality, suppose $\mu(S_1') \leq (1 - \Omega(1))\mu(S_1)$. Then
$$\int_{S_1}  P_x(S_2) d\mu(x) \geq  \Omega(\mu(S_1))$$ and we are
done.
\end{proof}

\begin{proof}[Proof of Lemma~\ref{1lem:l2}]
{ We remind the reader that 
we are given a convex set $K$ containing the origin as an interior point and a linear objective $c$, such that
  $$Q := K \cap \{y : c^Ty \leq
1\}$$ is bounded, for any chord $\overline{pq}$ of $Q$ passing through the origin, $\frac{|p|}{|q|} \leq s$    and $\eps
> 0$.  Also, $T: Q \ra \RR^n$ is defined by
$$T(x) = \frac{ x}{ 1 - c^T x}.$$ } 
Let $K_\eps$ be $K \cap \{c^T x \leq 1-\eps\}$ and $\tk_\eps := T K_\eps$.
Given four collinear points $a, b, c, d$,  $(a:b:c:d) :=
\frac{(a-c)\cdot(b-d)}{(a-d)\cdot(b-c)}$ is called the the cross
ratio.
Let $\overline{p' q'}\ni 0$ be a chord of $\tk_\eps$,
and $p = T^{-1}(p')$ and $q = T^{-1}(q')$. If $c^T p' \leq c^T q'$, then $\frac{|p'|}{|q'|} \leq \frac{|p|}{|q|} \leq s$.  On the other hand, if $c^T p' \geq c^T q'$,
let $r$ be the intersection of $\overline{pq}$ with { $\{x: c^T x = 1\}$.}
By the projective invariance of the cross ratio (see for example, Lemma 14 in \cite{KNsample})
$$ (\infty:0:p':q') = (r:0:p:q).$$ Therefore
\beqs \frac{|p'|}{|q'|} & = & \left(\frac{|p|}{|q|}\right)\left(\frac{|q-r|}{|p-r|}\right)\\
& \leq & \left(s\right)\left(\frac{s+1}{\eps}\right). \eeqs
Thus  \beq \label{1bound11}\sup_{\overline{p'q'} \ni 0} \ln \frac{|p'|}{|q'|} = O\left(\ln \frac{s}{\eps}\right),\eeq where the supremum is taken over all chords of $K$ containing the origin.
By (\ref{1bound11}) and Theorem~\ref{1tself} and Theorem~\ref{1thyp}, it follows that
$$  \sup_{h \in \tk_\eps} \ln \|h\| \leq O\left(\ln \left(\frac{s \nu}{\eps}\right)\right),$$ and therefore
\beq \label{1volbd} \ln \vol \tk_\eps \leq O\left(n \ln \left(\frac{s \nu}{\eps}\right)\right).\eeq
Let $\rho$ be a density supported on $\tk$ such that $$\int_{\tk_\eps} \rho(x) dx \geq \de.$$
Then, \beqs \int_{\tk} \rho(x)^2 dx & \geq & \int_{\tk_\eps} \rho(x)^2 dx\\
& \geq &  \frac{\de^2}{\vol \tk_\eps}\, {\text{by convexity of the squared function and Jensen's inequality}}\\
& \geq & \de^2 \exp\left(- O\left(n \ln \left(\frac{s \nu}{\eps}\right)\right)\right), \,\,\,\,\,\text{by (\ref{1volbd})}.\eeqs
\end{proof}

\begin{proof}[{ Proof of Lemma~\ref{cl:1}}]


Let  $\yy_h(w) := D^2 {F_h}(w)$ and $\yy_s = D^2 {F_s}(w)$. Then, $\yy(w) = n \yy_h + n^2 \yy_s.$
Next,
\beq\label{1m11}\Tr\yy_h(z) - \Tr\yy_h(0) = D\Tr\yy_h(0)[z] + \frac{D^2\Tr\yy_h(z')[z, z]}{2},\eeq for some $z' \in [0, z]$.


 {\beqs |D\Tr\yy_h(0)[z]| & \leq &  O\left(n \left( \sup_{\|v\| = 1} |D^3 {F_h}[v, v, v]|\right)\right)\,\,\,\text{\wpot}\\
 & \leq & O\left(n \left( \sup_{\|_hv\| = 1/\sqrt{n}} |D^3 {F_h}[v, v, v]|\right)\right)\,\,\,\text{\wpot}\\
 & \leq & O(1/\sqrt{n})\,\,\,\, \text{\wpot}.\eeqs }
Applying Lemma~\ref{1lnice}, we see that
 \beq \label{1l1111}\p[- D\Tr\yy_h(0)[z] < O(1/n)] > \frac{998}{1000}.\eeq
%
Next, 
{\beq \lab{(16)} D^2\Tr\yy(z')[z, z] & = &  D^2\Tr D^2{(nF_h + n^2 F_s)}(z')[z, z].\eeq
In order to bound  (\ref{(16)}),} let $A$ be an invertible matrix such that $\yy(z') = A^T\yy(0)A$. Such a matrix $A$ exists for which  $\|A - I \| = O(1/\sqrt{n})$ \wpot because $\|_h z'\|  = O(1/\sqrt{n})$ \wpot. Let $D_A$ be the differential operator whose action on a function $G$  is determined by the relation
$$\forall v\in \RR^n, \,D_A G(w)[v] := D G(w)[Av].$$ Thus $D_A^2 {F_h}(z') = \yy_h(0)$.
Now,
\beq \|_hu\| = O(1) \Rightarrow \forall \|_h v\| = 1, D^4{F_h}(z')[u, u, v, v] \leq O(1).\label{1eq347}\eeq

  \beq
D^2\left(\Tr\yy_h(z')\right)[z, z] & = &  D^2 \left(\Tr D_A^2 {F_h}(0)\right)[z, z] \nonumber\\
& \leq & n \sup_{\|Av\| = 1} O(D^4 F(z')[v, v, z, z]) \,\,\,\text{\wpot}\nonumber\\
& \leq  &  n^2 \sup_{\|_h v\| = 1/\sqrt{n}} D^4F(z') [v, v, v, v]\|_h z\|^2 \,\,\,\,\,\,(\text{by Fact~\ref{1f1}})\nonumber\\
& = & O(1/n) \,\,\,\,\,\,\, \text{\wpot}. \lab{1l2222}\eeq
The last line here uses Lemma~\ref{lem:1}.

Therefore, by Equations~\ref{1m11}, \ref{1l1111} and \ref{1l2222}, we have
$$\p\left[  - \left(\Tr\yy_h(z) - \Tr\yy_h(0)\right) < O(1/n) \right] > \frac{997}{1000}.$$
{Also \wpot, $\|_sz\| = O(1/n)$.} Therefore \wpot,
\beq\label{1m1129}\Tr\yy_s(z) - \Tr\yy_s(0) & = & O(\Tr\yy_s(0)\|_sz\|)\\
& = & O(1/n^2). \eeq
The statement follows from the last two sentences, since $\yy = n \yy_h + n^2 \yy_s.$

\end{proof}

\begin{proof}[Proof of Lemma~\ref{cl:2}]{\beq\label{1m11new}-\Tr\xx(z) + \Tr\xx(0) =  -\left(D\Tr\xx(0)[z] + \frac{D^2\Tr\xx(z')[z, z]}{2}\right),\eeq for some $z' \in [0, z]$.}
Lemma 12  in  \cite{KNsample} shows that  $\|_\ell \nabla \Tr \xx\| \leq 2 \sqrt{n}$. Since for all vectors $v$, $\| v \| \geq \|_\ell v\|,$ this implies that   $\|\nabla  \Tr \xx\| \leq 2 \sqrt{n}$.
{ By Lemma~\ref{1lnice}, this implies that \beqs \p\left[|D\Tr\xx(0)[z]| < O(1)  \right] \geq 1 - 10^{-3}.\eeqs
By Lemma 13 in \cite{KNsample}, $\frac{- D^2\Tr\xx(z')}{2}\leq 0,$ thereby completing the proof.}
\end{proof}

\begin{proof}[Proof of Lemma~\ref{cl:2.5}] { Recall that $W(0) + Z(0) = I$.}

In order to prove that \beqs \p\left[\big|\Tr\left(I - (\xx(z) + \yy(z))\right)^3\big| \leq O(1)\right] \geq \frac{99}{100}, \eeqs
it suffices to show that \beq \lab{eq:c3-1}\p\left[ \| (\xx(z) - \xx(0)) \| \leq O(n^{-1/3})\right] \geq 1 - 10^{-3}\eeq and that
{ \beq \lab{eq:c3-2}\p\left[\| (\yy(z) - \yy(0))\| \leq O(n^{-1/2})\right] \geq 1 - 10^{-3}.\eeq}
From Lemma~\ref{2lnice} and Fact~\ref{thm:NT2}, we obtain (\ref{eq:c3-1}).
We obtain (\ref{eq:c3-2}) from (\ref{eq:dikinfund}).

{ In order to prove that \beqs \p\left[\big|\Tr\left(I - (\xx(z) + \yy(z))\right)^2\big| \leq O(1)\right] \geq \frac{99}{100}, \eeqs
it suffices to show that \beq \lab{c2.5-1}\p\left[ \Tr (\xx(z) - \xx(0))^2 \| \leq O(1)\right] \geq 1 - 10^{-3}\eeq and that

\beq \lab{eq:c2.5-2}\p\left[\Tr (\yy(z) - \yy(0))^2 \leq O(1)\right] \geq 1 - 10^{-3},\eeq
since by Cauchy-Schwartz,
$$|\Tr (\xx(z) - \xx(0))(\yy(z)-\yy(0))|^2 \leq \Tr\left( (\xx(z) - \xx(0))^2\right) \Tr \left((\yy(z) - \yy(0))^2\right) .$$
The above inequality (\ref{eq:c2.5-2}) follows from (\ref{eq:c3-2}).
We will prove (\ref{c2.5-1}) below. We have 
\begin{eqnarray*}
\Tr\left((W(z) - W(0))^2\right) & = & \Tr\left(\sum_{i \leq m} a_i a_i^T\left(\frac{1}{(1 - a_i^Tz)^2} - 1\right)\right)^2\\
 & \leq & \Tr\left(\sum_{i \leq m} a_i a_i^T ((2a_i^T z) + O(|a_i^Tz|^2))\right)^2\\
& \leq & O\left(\Tr\left(\sum_i \sum_j  (a_i a_i^T)(a_j a_j^T)((2a_i^T z) (2a_j^T z) + O(n^{-\frac{5}{4}})) \right)\right)\\
& \leq & \sum_i \sum_j (a_i a_j^T)^2  ((2a_i^T z) (2a_j^T z) + O(n^{-\frac{5}{4}})).
\end{eqnarray*}
(The last three lines above are true with probability $> 1 - 10^{-4}$.)
We proceed to obtain a bound on $$\E \sum_i \sum_j (a_i a_j^T)^2  ((2a_i^T z) (2a_j^T z) + O(n^{-\frac{5}{4}}))$$ and then the lemma follows from Markov's inequality.

$$\E \sum_i \sum_j (a_i a_j^T)^2  ((2a_i^T z) (2a_j^T z) + O(n^{-\frac{5}{4}}))  $$ is less or equal to 
\begin{eqnarray*}
 \sum_i \sum_j (a_i a_j^T)^2 (\sqrt{\E((a_i^Tz)^2)\E((a_j^Tz)^2)} + O(n^{-\frac{5}{4}}))
& \leq & \sum_i \sum_j (a_i a_j^T)^2 (O(1/n))\\
& = & O(1/n) \Tr\sum_i  a_ia_i^T \sum_j a_j a_j^T\\
& = &  O(1).
\end{eqnarray*}}
\end{proof}


\begin{proof}[Proof of Lemma~\ref{l:ell}]

 Fixing an orthonormal basis with respect to

 $< \cdot, \cdot>$,
$$\sum_{i = 1}^m a_i a_i^T \preceq I,$$ where $X  \preceq Y$ signifies that $Y$ dominates $X$ in the semidefinite cone.

Recall that for any $v$ such that $\|v\|= 1$, $\E(<v, z>^2) = r^2 < 1/C$ for some sufficiently large constant $C$.
It suffices to prove the following two inequalities.
\begin{lemma}\lab{lem:lat1}
\beqs\p\left[\max\left(\|_\ell z\|_y^2 - \|_\ell z\|^2, <_\ell  y, z>_y,   <_\ell  y, z>_z \right)  < O(\frac{1}{n}) \right] & > & \frac{19}{20}.\label{ell1}\eeqs
\end{lemma}
\begin{lemma}\lab{lem:lat2}
\beqs\p\left[ \|_\ell z \|_z^2 - \|_\ell z\|^2 < O(\frac{1}{n}) \right] & > & \frac{9}{20}.\label{ell2}\eeqs
\end{lemma}
\end{proof}

\begin{proof}[Proof of Lemma~\ref{l:h}]
We will prove upper bounds on each of  (a) $\|_h z\|_y^2 - \|_h z\|^2,$ (b) $<y, z>_y,$ (c) $\|_h z \|_z^2 - \|_h z\|^2$ and (d) $<_h y, z>_z$ that hold with constant probability, and then use  the union bound.
We will repeatedly use the observation (that holds from Fact~\ref{f:s1}) that for any point $w$ such that $\|_h w\| = o(1)$,
\beq \|_h y\|_w \leq O(n^{-1/2} \|y\|) \leq O\left(\frac{1}{n}\right)\lab{eq:39h}\eeq and \wpot \beq \|_h z\|_w \leq O(n^{-1/2} \|z\|) \leq O(1/\sqrt{n})\lab{eq:40h}.\eeq
\ben
\item[(a)] \beqs \label{1e12h}\|_h z\|_y^2 - \|_h z\|^2 & = & D^2{F_h}(y)[z, z] - D^2{F_h}(0)[z, z]\\
& = & D^3{F_h}[w][y, z, z],\eeqs
for some $w$ on the line segment $[0, y]$.

By Fact~\ref{1f1},
\beqs \p\left[D^3{F_h}(w)[y, z, z] < O(n^{-2})\right]
& \geq & \p\left[ \|_h y\|_{w} (\|_h z\|_{w})^2  \leq  O(\frac{1}{n^2})\right].\eeqs
Since $\| z\|^2 = \|_\ell z \|^2 + n \|_h z\|^2 + n^2 \|_s z\|^2,$ we have
 $$ \|_h z\|_{w}  =   O(\|z\|_{w}/\sqrt{n}).$$
 Also, $\|_h w\|  = O(\|y\|/\sqrt{n}) = O(1/n),$ and so \wpot,
 $$O(\|_hz\|_{w}) = O(\|z\|/\sqrt{n}) = O(1/\sqrt{n}).$$  Thus,
\beq \lab{e:30}\p\left[ \|_h y\|_{w} \|_h z\|_{w}  \leq  O(\frac{1}{n\sqrt{n}})\right] & \geq & \ot. \eeq and
 \beq \lab{e:31}\p\left[ \|_h y\|_{w} (\|_h z\|_{w})^2  \leq  O(\frac{1}{n^2})\right] & \geq & \ot. \eeq

\item[(b)] \beq \nonumber <_h y, z>_y & = & <_h y, z> + (<_h y, z>_y - <_h y, z>)\nonumber\\
& = & <_h y, z> + D^3{F_h} (w)[y, y, z]             \,\,\,\,\,\,\text{(for  $w \in [0, y]$)} \lab{e:33h}\\
& = & O(\frac{\|_h y\|}{{n}}) + O(\|_h y\|_{w}^2\|_h z\|_{w}) \,\,\,\,\text{with probability} > 99/100. \lab{e:34h}\eeq
In going from (\ref{e:33h}) to (\ref{e:34h}), we used Fact~\ref{1f1} and Fact~\ref{1f2}. In the above calculation, to ensure that $w$ is well-defined, we take it to be the candidate with the least norm.
{ Thus by Equations (\ref{e:30}), (\ref{e:31}) and (\ref{e:34h}),}
{ \beqs \p\left[  (<_h y, z>_y) < O(\frac{1}{n}) \right] & \geq & \p\left[\frac{\|_h y\| \|_h z\|}{\sqrt{n}} + \|_h y\|_{w}^2\|_h z\|_{w} = O\left(\frac{1}{n}\right)\right]\\
&  > & 98/100. \nonumber\eeqs}

\item[(c)] For some $w \in [0, z]$, \beq \,\,\,\,\,\,\,\,\,\,\,\,\,|\|_h z \|_z^2 - \|_h z\|^2 - D^3{F_h}(0)[z, z, z]| \leq \\\sup\limits_{w \in [0, y]} \big |\frac{D^4{F_h}(w)[z, z, z, z]}{2} \big |\lab{eq:itemc}\eeq
    By Lemma~\ref{1lnice}, $$\p\left[ D^3{F_h}(0)[z, z, z]  = O\left(\frac{\sup_{\|_h v\| \leq 1} D^3 {F_h} (0)[v, v, v]\|_h z\|^3}{\sqrt{n}}\right)\right] > 99/100.$$
     $\p[ \|_h z\| = O(1/\sqrt{n})] > 99/100,$
     therefore, each term in (\ref{eq:itemc}) is $O(1/n^2)$ with probability $\frac{99}{100}$ by Lemma~\ref{lem:1}, and so
\beqs\p\left[ \|_h z \|_z^2 - \|_h z\|^2 < O(\frac{1}{n^2}) \right] & > & \frac{98}{100}.\eeqs
\item[(d)] { \beqs <_h y, z>_z & = & <_h y, z> + (<_h y, z>_z - <_h y, z>)\\
& \leq & <_h y, z> + \sup_{w \in [0, z]}  \big | D^3{F_h}(w)[y, z, z] \big |            \\
& = & O(\frac{\|_h y\| }{n}) + 2\|_h y\|_{w}\|_h z\|_{w}^2 \,\,\,\,\text{with probability} > 99/100. \eeqs}
{ By Equations (\ref{eq:39h}) and (\ref{eq:40h}),}
$$\p\left[\frac{\|_h y\| }{{n}} + \|_h y\|_{w}\|_h z\|_{w}^2  = O\left(\frac{1}{n^2}\right)\right] > 99/100.$$
Therefore,
\beqs\p\left[(<_h y, z>_z)  < O(\frac{1}{n^2}) \right] & > & \frac{98}{100}.\eeqs
\een
\end{proof}
\begin{proof}[Proof of Lemma~\ref{l:s}]
We trace the same steps involved in the proof of the last lemma, the only difference being that of scale. We proceed to prove   upper bounds of $O(1/n^3)$ on each of the terms (a) $\|_s z\|_y^2 - \|_s z\|^2$ (b) $<_s y, z>_y,$ (c) $\|_sz \|_z^2 - \|z\|^2$ and (d) $<_s y, z>_z$ that hold with constant probability separately, and then use the union bound. We will repeatedly use the observation (that holds from Fact~\ref{f:s1}) that for any point $w$ such that $\|_s w\| = o(1)$,
\beq \|_s y\|_w \leq O(n^{-1} \|y\|) \leq O\left(\frac{1}{n\sqrt{n}}\right)\lab{eq:39s}\eeq and \wpot \beq \|_s z\|_w \leq O(n^{-1} \|z\|) \leq O(1/n)\lab{eq:40s}.\eeq
\ben
\item[(a)] \beqs \label{1e123s}\|_s z\|_y^2 - \|_s z\|^2 & = & D^2{F_s}(y)[ z, z] - D^2{F_s}(0)[z, z]\\
& = & D^3{F_s}(w)[y, z, z],\eeqs
for some $w$ on the line segment $[0, y]$.

\beqs  D^3{F_s}(w)[y, z, z] & \leq &  \|_s y\|_{w} \|_s z\|_{w}^2  \,\,\,\,\,\,\,\,\text{\byfactone}\\
& \leq &  (n^{-3/2})(n^{-1})^2 \,\,\,\,\,\,\,\,\text{\wp $ > 1 - 10^{-3}$}\eeqs

\item[(b)] \beq \nonumber <_s y, z>_y & = & <_s y, z> + (<_s y, z>_y - <_s y, z>)\\
& = & <_s y, z> + D^3{F_s}(w)[y, y, z]             \lab{e:33s}\\
& = & O\left(\frac{\|_s y\|}{n \sqrt{n}}\right) + O\left(\|y\|_{w}^2\|z\|_{w}\right) \,\,\,\,\text{with probability} > 99/100. \lab{e:34s}\eeq
In going from (\ref{e:33s}) to (\ref{e:34s}), we used Fact~\ref{1f1} and Fact~\ref{1f2}. We see that
{ $$\p\left[\frac{\|_s y\|}{n \sqrt{n}} + 2\|_s y\|_{w}^2\|_s z\|_{w} = O\left(\frac{1}{n^3}\right)\right] > 99/100,$$}
Therefore,
\beqs\p\left[(<_s y, z>_y) < O(\frac{1}{n^3}) \right] & > & \frac{98}{100}.\eeqs

\item[(c)] { \beqs \|_s z\|_z^2 - \|_s z\|^2 & = & D^2{F_s}(z)[ z, z] - D^2{F_s}(0)[z, z]\\
& = & D^3{F_s}(w)[z, z, z]\eeqs}
for some $w$ on the line segment $[0, z]$.
By Fact~\ref{1f1}, { \beqs D^3{F_s}(w)[z, z, z] & \leq &  2\|_sz\|_{w}\|_sz\|_w^2\\
& \leq & O\left(\frac{1}{n^3}\right) \,\,\,\,\,\text{\wpot}. \eeqs}

\item[(d)]  { \beqs <_s y, z>_z & = & <_s y, z> + (<_s y, z>_z - <_s y, z>)\\
& \leq & <_s y, z> + \sup_{w \in [0, z]} D^3{F_s}(w)[y, z, z]   \\
& = & O(\frac{\|_s y\| }{n\sqrt{n}}) + 2\|_s y\|_{w}\|_s z\|_{w}^2 \,\,\,\,\text{with probability} > 99/100. \eeqs}
By Equations (\ref{eq:39s}) and (\ref{eq:40s}),
{ $$\p\left[\frac{\|_s y\| \|_s z\|}{\sqrt{n}} + 2\|_s y\|_{w}\|_s z\|_{w}^2  = O\left(\frac{1}{n^3}\right)\right] > 99/100.$$}
Therefore,
\beqs\p\left[(<_s y, z>_z)  < O(\frac{1}{n^3}) \right] & > & \frac{98}{100}.\eeqs
\een
\end{proof}

\begin{proof}[Proof of Lemma~\ref{lem:lat1}]
\begin{enumerate}
\item[(a)] {\beqs \label{1e12l}\|_\ell z\|_y^2 - \|_\ell z\|^2 & = & D^2{F_\ell}(y)[z, z] - D^2{F_\ell}(0)[z, z]\\
& = & D^3{F_\ell}(w)[y, z, z], \eeqs}
for some $w \in [0, y]$  and consequently $\|_\ell w\| = O(1/\sqrt{n})$ and hence
{ \beqs   D^3{F_\ell}(w)[y, z, z] = D^3{F_\ell}(0)[y, z, z] + D^4{F_\ell}(w_1)[y, y, z, z].\eeqs
The term $D^4{F_\ell}(w_1)[y, y, z, z]$ is bounded above $O(1/n)$ with probability $1 - 10^{-3}$. We apply Cauchy-Schwartz below.
\beq  \E\left[ \left(\sum_{i=1}^m (y^T a_i a_i^T z)(a_i^T z) \right)^2\right] & \leq & \E \left(\sum_{i=1}^m (y^T a_i a_i^T z)^2\sum_i (a_i^T z)^2 \right)\nonumber\\
& \leq & \E\left(\sum_i\|( a_i a_i^T)y\|^2\| z\|^4/n\right)\nonumber\\
& = & O(1/n^2),\,\,\,\,\,\text{since $\|\sum_{i}  (a_i a_i^T)^2\| \leq 1.$} \eeq
Therefore, \beq \nonumber  \p\left[D^3{F_\ell}(0)[y, z, z] < O(1/{n})\right] & = &  \p\left[ 2 \sum_{i=1}^m {(a_i^T y)(a_i^T z)^2} < O(1/{n})\right]\\
& \geq & 1 - O\left(n^2 \E\left[ \left(\sum_{i=1}^m (y^T a_i a_i^T z)(a_i^T z)\right)^2\right]\right)\\
& \geq & 1 - 10^{-3} \,\,\,\,\text{rescaling $y$ by a universal constant}.\lab{eq:us2}\eeq}
\item[(b)]
Proceeding to the next term, 
\beqs \p\left[ <_\ell  y, z>_y = O(1/{n})\right] & = &\p\left[ <_\ell  y, z> + (<_\ell  y, z>_y - <_\ell  y, z>) = O(1/{n})\right].\nonumber\eeqs
{ Also,} $$<_\ell  y, z>_y - <_\ell  y, z> = O(\|z\|/n)$$ by (\ref{eq:dikinfund}), and
\beq \lab{eq:us1}\E[ <_\ell  y, z>^2] \leq O(1/n^2), \eeq { so by Markov's inequality}
we obtain
 \beq \p\left[ <_\ell  y, z>_y = O(1/n)\right] > 1 - 10^{-3}.\lab{e:need1.5}\eeq
\item[(c)]
 { Finally, we obtain a probabilistic upper bound  \beq \p[<_\ell  y, z>_z - <_\ell  y, z> < O(1/n)] > 1 - 10^{-3},  \lab{eq:lntr}\eeq
as follows.
Note that \beq<_\ell  y, z>_z - <_\ell  y, z> =  y^T \sum_i \left(\frac{a_i a_i^T z}{(1-a_i^Tz)^2} - a_i^Tz\right). \eeq
This is equal to 
\beqs \sum_i y a_i^T\left(\frac{2 (a_i^Tz)^2 - (a_i^Tz)^3}{(1 - a_i^T z)^2}\right).\eeqs

Let the $a_i$ be listed in order of non-increasing length. For $i=1$  to $n$, the probability that $|a_i^T z| \geq
n^{-\frac{1}{4}}$ is $O(e^{-\sqrt{n}/2})$. For $i >n$,  $|a_i^T z| \geq
n^{-\frac{1}{4}}$ is $O(e^{-\sqrt{i}/2})$. It is true with probability $1-10^{-3}$ that $\forall_i |a_i^T z|$ is $ \leq
\|a_i^T\| r$, which is less than $\frac{1}{2}$.
Therefore with probability $> 1 - 10^{-3}$ every term $\frac{2 (a_i^Tz)^2 - (a_i^Tz)^3}{(1 - a_i^T z)^2}$ is bounded above by $3 (a_i^Tz)^2$.

$$\sum_i y a_i^T\left(\frac{2 (a_i^Tz)^2 - (a_i^Tz)^3}{(1 - a_i^T z)^2}\right)  < 
 \|y\|\left\|\sum_i  a_i\left(\frac{2 (a_i^Tz)^2 - (a_i^Tz)^3}{(1 - a_i^T z)^2}\right)\right\| $$

Next, we apply the Semidefinite Cauchy-Schwartz inequality from \cite{KNsample} and take operator norms on both sides. With probability $> 1 - 10^{-3}$, 

$$\left\|\sum_i  a_i\left(\frac{2 (a_i^Tz)^2 - (a_i^Tz)^3}{(1 - a_i^T z)^2}\right)\right\| ^2 \leq \|\sum_i a_i a_i^T\| \sum_i 9 (a_i^Tz)^4 \leq O\left(\sum_i (a_i^Tz)^4\right).$$ 
Next
$$\E[\sum_i (a_i^Tz)^4] = O\left(\frac{\sum_i\|a_i\|^4}{n^2}\right) = O\left(\frac{1}{n}\right).$$ The last step uses the fact that each $\|a_i\| \leq 1$ and $\sum_i \|a_i\|^2 \leq n$.
Thus $$\p\left[\sum_i y a_i^T\left(\frac{2 (a_i^Tz)^2 - (a_i^Tz)^3}{(1 - a_i^T z)^2}\right) < O(1/n)\right] > 1 - 10^{-3}.$$}

{ It follows that
\beq \p\left[<_\ell  y, z>_z - <_\ell y, z> = O(1/{n})\right] > 1 - 10^{-3}.\lab{e:need2}\eeq}
\end{enumerate}
{ Lemma~\ref{lem:lat1} follows.}
\end{proof}
\begin{proof}[Proof of Lemma~\ref{lem:lat2}]
In order to prove that
\beqs\p\left[ \|_\ell z \|_z^2 - \|_\ell z\|^2 < O(\frac{1}{n}) \right] & > & \frac{9}{20},\eeqs
it suffices to show that
\beqs\p\left[ \left(\|_\ell z \|_z^2  + \|_\ell z \|_{-z}^2 \right)/2 <  \|_\ell z\|^2 + O(\frac{1}{n}) \right] & > & \frac{9}{10},\eeqs because the distribution of $z$ is symmetric about the origin.
\beq\nonumber \sum_i \left(\frac{(a_i^T z)^2}{2(1- a_i^T z)^2} +
\frac{(a_i^T z)^2}{2(1 + a_i^T z)^2} \right) & = & \sum_i
(a_i^Tz)^2\left(\frac{1 + (a_i^Tz)^2}{(1 - (a_i^Tz)^2)^2}\right)\\ &
= &
 \sum_i \left((a_i^Tz)^2 + \frac{3(a_i^Tz)^4 - (a_i^T z)^6}{(1- (a_i^Tz)^2)^2}\right)\nonumber\\ & = &
\|_\ell z\|^2 +  \sum_i \frac{3(a_i^Tz)^4 - (a_i^T z)^6}{(1-
(a_i^Tz)^2)^2} \label{eq:bas1}.\eeq

{ Let the $a_i$ be listed in order of non-increasing length. For $i=1$  to $n$, the probability that $|a_i^T z| \geq
n^{-\frac{1}{4}}$ is $O(e^{-\sqrt{n}/2})$. For $i >n$,  $|a_i^T z| \geq
n^{-\frac{1}{4}}$ is $O(e^{-\sqrt{i}/2})$. It is true with probability $1-10^{-3}$ that $\forall_i |a_i^T z|$ is $ \leq
\|a_i^T\| r$, which is less than $\frac{1}{2}$. This allows us to
write \wpot \beqs \E\left[\frac{3(a_i^Tz)^4 - (a_i^T z)^6}{(1-
(a_i^Tz)^2)^2}\right] & = & 3\E[(a_i^Tz)^4](1+o(1)),\eeqs}
which is $O(\|_\ell a_i\|^4/n^2)$. Since
$\sum_i a_i a_i ^T \preceq I,$ therefore,
$\forall i,\,\|_\ell a_i\|\leq 1$ and $\sum_i \|_\ell a_i\|^2 \leq n.$

Therefore,
{ \beqs  \p\left[ \left(\|_\ell z \|_z^2  + \|_\ell z \|_{-z}^2 \right)/2  >  \|_\ell z\|^2 + O\left(\frac{100}{n}\right) \right]  & = & \p\left[\sum_i \frac{3(a_i^Tz)^4 - (a_i^T z)^6}{(1-
(a_i^Tz)^2)^2} > O\left(10^2/n\right)\right]\\ & \leq & n \sum_i  \|_\ell a_i\|^4/(100 n^2)\\
& \leq & \sum_i \|_\ell a_i\|^2/(100 n)\\
& \leq & 1/100. \eeqs}
\end{proof}


\end{document}